\documentclass[a4paper, dvipsnames, 11pt]{article}
\usepackage[utf8]{inputenc}
\usepackage{jcappub}
\usepackage{amsfonts}
\usepackage{graphicx}
\usepackage{amssymb}
\usepackage{amsmath}
\usepackage{breqn}
\usepackage{array}
\usepackage{float}
\usepackage{tikz}
\usepackage{subfig}
\usepackage{xcolor}
\usepackage{multirow}
\usepackage{afterpage}
\usetikzlibrary{shapes.geometric, arrows}
\tikzstyle{st} = [rectangle, rounded corners, text width = 3cm, text centered, draw = black ]
\tikzstyle{arrow} = [->,>=stealth]

\newcommand{\g}{\gamma}

\newcommand{\viz}{\textit{viz.~}}

\title{Sub-Horizon Amplification of Curvature Perturbations: A New Route to Primordial Black Holes and Gravitational Waves}
\author[a, b]{Debottam Nandi,}
\author[b]{Rohan Roy,}
\author[b]{Simran Yadav,}
\author[b]{and Arnab Sarkar}
\affiliation[a]{Center for Cosmology and Science Popularization, SGT University, Gurugram, Haryana 122505, India}
\affiliation[b]{Department of Physics and Astrophysics, University of Delhi, Delhi 110007, India}

\emailAdd{debottam\_ccsp@sgtuniversity.org}

\emailAdd{rohankgp.2021@gmail.com}

\emailAdd{simranyadavkhola@gmail.com}

\emailAdd{arnabsarkar555444@gmail.com}

\abstract{
   The enhanced primordial scalar power spectrum is a widely studied mechanism for generating primordial gravitational waves (PGWs), also referred to as scalar-induced gravitational waves (SIGWs). This process also plays a pivotal role in facilitating the formation of primordial black holes (PBHs). Traditionally, the ultra slow-roll (USR) mechanism has been the predominant approach used in the early universe. In this framework, the second slow-roll parameter $\epsilon_2$, is typically set to $-6$ or lower for a brief period --- marking a significant departure from the standard slow-roll condition where $\epsilon_2 \simeq 0$. Such conditions often emerge in models with inflection points or localized features, such as bumps in the potential. In this paper, we challenge the conventional assumption that $\epsilon_2 \lesssim -6$ is a prerequisite for substantial amplification of the scalar power spectrum. We demonstrate that any negative value of the second slow-roll parameter can indeed enhance the scalar power spectrum through sub-horizon growth, establishing this as a necessary and sufficient condition for amplification. Consequently, this mechanism facilitates the generation of both PGWs and PBHs. To illustrate this, we examine a standard scenario where a brief USR phase is embedded between two slow-roll (SR) phases. By systematically varying $\epsilon_{2}$ values from $-1$ to $-10$ in the USR region, we investigate the amplification of the power spectrum and its implications for PGWs and PBHs production, particularly in the context of ongoing and future cosmological missions.
}

\begin{document}

\maketitle

\section{Introduction}
To date, Cosmic Microwave Background Radiation (CMBR)~\cite{Planck:2018jri, Planck:2018vyg} has been at the forefront of probing the dynamics of the early Universe, with the inflationary paradigm \cite{Starobinsky:1979ty, Starobinsky:1980te, Guth:1981, Mukhanov:1981xt, LINDE1982389, HAWKING1982295, Guth:1982, Sasaki1986, Albrecht-Steinhardt:1982, Linde:1983gd, VILENKIN1983527, Bardeen:1983, Mukhanov:1990me, Liddle:1994dx, Lidsey:1995np,  Copeland:1997et, Lyth:2007qh, Sriramkumar:2009kg, Martin:2013tda, Martin:2015dha} characterized by a rapid acceleration of the Universe, successfully explaining the observation. However, since the initial detection of gravitational waves (GWs) by LIGO-VIRGO collaborations in 2015 \cite{LIGOScientific:2016aoc, LIGOScientific:2016emj, LIGOScientific:2016jlg, LIGOScientific:2016sjg, LIGOScientific:2016vbw, LIGOScientific:2016vlm, LIGOScientific:2017bnn, LIGOScientific:2017vox, LIGOScientific:2017vwq, LIGOScientific:2017ycc, LIGOScientific:2018glc, LIGOScientific:2020aai, LIGOScientific:2020stg, LIGOScientific:2020zkf}, a new method to investigate the early Universe has emerged. This has been further bolstered by recent detections of nano-Hertz GWs by NANOGrav \cite{NANOGrav:2020bcs, NANOGrav:2021flc, NANOGrav:2023ctt, NANOGrav:2023gor, NANOGrav:2023hde, NANOGrav:2023hvm, NANOGrav:2023icp, NANOGrav:2023pdq, NANOGrav:2023tcn, NANOGrav:2023hfp} and other Pulsar Timing Array (PTA) collaborations \cite{Goncharov:2021oub, EPTA:2021crs, EPTA:2023akd, EPTA:2023fyk, EPTA:2023gyr, EPTA:2023sfo, EPTA:2023xxk, EuropeanPulsarTimingArray:2023egv, Antoniadis:2022pcn}, which strongly support the existence of a stochastic GW background and thus open promising avenues for exploring early Universe physics \cite{Assadullahi:2009jc, Gong:2015qha, Bartolo:2016ami, Cui:2018rwi, Domenech:2020kqm, Liu:2023ymk}. While numerous sources such as cosmological first-order phase transitions \cite{Nakai:2020oit, Freese:2022qrl, Morgante:2022zvc, Bringmann:2023opz}, annihilating domain wall networks \cite{Wang:2022rjz, Ferreira:2022zzo}, cosmic strings \cite{Blasi:2020mfx, Ellis:2020ena, Buchmuller:2020lbh}, supermassive black hole binaries (SMBHs)\cite{Bean:2002kx, Duechting:2004dk, Shannon:2015ect, Pandey:2018jun, Becsy:2022pnr, Ellis:2023owy, Ellis:2023dgf}, axion dynamics \cite{Unal:2023srk, Lozanov:2023rcd} could explain these observations, GWs generated during inflation are a particularly promising candidate \cite{Bartolo:2001cw, Bartolo:2004if, Inomata:2018epa, Inomata:2023zup}.

There are two primary mechanisms by which GWs can be produced during the early Universe. The First mechanism involves the amplification of the GWs due to the linear dynamics of the tensor perturbations \cite{Bruni:1996im, Carbone:2004iv, Ananda:2006af, doi:10.1142/9789814327183_0010, Gao:2012ib, Pi:2019ihn, Akama:2019qeh, Zhu:2021whu, Galloni:2022mok}. On the other hand, the second, and the most promising mechanism occurs when the linear tensor perturbations are amplified by the higher-order amplified scalar perturbations \cite{Matarrese:1997ay, Acquaviva:2002ud, Bartolo:2003bz, Vernizzi:2004nc, Nakamura:2004rm, PhysRevD.69.104011, Inomata:2021tpx, Baumann:2022mni}, known as the scalar-induced gravitational waves (SIGWs)~\cite{Domenech:2021ztg, Dandoy:2023jot}. As a consequence, large scalar fluctuations can later collapse and form primordial black holes (PBHs) \cite{Kawasaki:1997ju, Yokoyama:1998pt, Kawasaki:1998vx, Rubin:2001yw, Khlopov:2004sc, Saito:2008em, Khlopov:2008qy, Drees:2011yz, Drees:2011hb, Ezquiaga:2017fvi, Kannike:2017bxn, Hertzberg:2017dkh, Pi:2017gih, Dalianis:2018frf, Cicoli:2018asa, Ozsoy:2018flq, Byrnes:2018txb, Ballesteros:2018wlw, Belotsky:2018wph, Martin:2019nuw, Ezquiaga:2019ftu, Fu:2019ttf, Kawai:2021edk, Escriva:2022duf, Kristiano:2022maq, Riotto:2023hoz, Ozsoy:2023ryl, Choudhury:2024aji, Choudhury_2024,Carr:1975qj, Carr:2009jm, Carr:2016drx, Carr:2018rid, Sasaki:2018dmp, Hawking:1975vcx, Carr:1974nx, Chapline:1975ojl, Yokoyama:1998qw, Niemeyer:1997mt, Musco:2004ak, Carr:2020gox, Garcia-Bellido:2017fdg, Carr:2023tpt} when they re-enter the horizon, which are also considered a promising candidate for explaining present abundances of dark matter (DM) \cite{Ivanov:1994pa, Alcock:1998fx, Allsman:2000kg, Capela:2012jz, Capela:2013yf, Griest:2013esa, Graham:2015apa, Clesse:2015wea, Clesse:2016vqa, Bird:2016dcv, Chen:2016pud, Blum:2016cjs, Urena-Lopez:2016yon, Ballesteros:2017fsr, Garcia-Bellido:2017aan, Inomata:2017okj, PhysRevLett.111.181302, Bartolo:2018evs, Bartolo:2018rku, Katz:2018zrn, Maity:2018exj, Bhaumik:2019tvl, Montero-Camacho:2019jte, Dasgupta:2019cae, Laha:2019ssq, Laha:2020ivk, Green:2020jor, Jedamzik:2020omx, Jedamzik:2020ypm, Carr:2020xqk, DeLuca:2020agl}.

There are various methods to generate such enhanced curvature perturbation \cite{Animali:2022otk, Ezquiaga:2017fvi, Motohashi:2017kbs, Cai:2021zsp, Boutivas:2022qtl, Choudhury:2024aji, Aldabergenov:2020yok, Kanazawa:2000ea, Kamenshchik:2018sig, Gao:2020tsa, Lin:2020goi, Ballesteros:2017fsr, Leach:2000ea, Yokoyama:1998pt, Balaji:2023ehk, Gorji:2023sil, Garcia-Bellido:2017mdw, Heydari:2024bxj, Ragavendra:2020sop,Arya:2019wck, Bastero-Gil:2021fac, Sugimura_2012, Sasaki_2008}. Among them, the single canonical scalar field minimally coupled to the gravity is the most significant and yet the successful one \cite{Garcia-Bellido:2017mdw, Germani:2017bcs, Kefala:2020xsx, Inomata:2021uqj, Inomata:2021tpx, Choudhury:2024aji,Mishra:2019pzq, ZhengRuiFeng:2021zoz,Dimopoulos:2017ged, Wu:2021mwy, Choudhury:2024aji, Kristiano:2024ngc, Gow:2020bzo}, which includes bump, valley or inflection point in the base potential. It's important to note that at the pivot scale, this model must adhere to the constraints set by the CMB. As a result, the amplification of the scalar spectrum must occur after this pivot scale. Additionally, this enhancement period should not be too prolonged to the extent that the perturbations become large enough to backreact with the background. To accommodate these requirements, we can divide the early universe into at least three distinct regimes. The first regime, which includes the regime where pivot scale leave the horizon, is characterized by standard slow-roll (SR-I), where both the slow-roll parameters are small compared to unity, i.e., $|\epsilon_1|,~ |\epsilon_2| \ll 1.$ This allows the primordial curvature perturbation to align well with the CMB constraints . However, the standard SR conditions are inadequate for sufficiently enhancing scalar perturbations which is required for SIGW as well as for production of PBHs. Therefore, following the SR regime, a second epoch of amplification of the curvature perturbations is considered to last for a brief period. In this case, we conventionally follow the standard way of amplifying the spectrum: the ultra slow-roll (USR), where the second slow-roll parameter $\epsilon_{2}$ violates the SR conditions and becomes $\lesssim -6.$ After the brief USR phase, again the conventional slow-roll (SR-II) resumes until inflation ends. In summary, the entire inflationary phase can be effectively understood by dividing it into three regions: the first slow-roll (SR-I), followed by ultra-slow roll (USR), and then the second slow-roll (SR-II).

In this article, we confine ourselves to the above straightforward scenario: single canonical scalar field minimally coupled with gravity, undergoing the conventional SR-USR-SR phases. Our primary objective is to address a fundamental question: {\bf  is the conventional USR regime, characterized  by $\epsilon^{\rm usr}_{2} \lesssim -6$ essential for enhancing the scalar spectrum? If not, what broader conditions facilitate the enhancement of scalar perturbations?} By analyzing the standard setup, we observe that near-zero derivatives of the potential, such as a flat region, inflection point or a bump in the potential necessitate a deviation from slow-roll, which consequently forces $\epsilon_{2}$  to approach $-6$ and results in $\epsilon^{\rm usr}_1 \propto a^{-6}.$ This leads to the amplification of the curvature perturbation in the super-horizon sale --- a phenomena traditionally identified as the conventional USR. However, our findings reveal that {\bf this is merely one instance within a broader set of criteria that can facilitate the enhancement.} Specifically, we identify that the key to enhancement lies not only in super-horizon amplification but rather also in sub-horizon conditions. We demonstrate that the sub-horizon scalar power spectrum is inversely proportional to $\epsilon_1,$ thus, any moderate attenuation in the first slow-roll parameter guarantees amplification in the power spectrum. Specifically, if the attenuated value of $\epsilon_1$ is denoted as $\alpha$ compared to its counterpart during the SR-I regime, the power spectrum of scalar perturbation will amplify correspondingly, as represented by the relation:

$$\epsilon_1^{\rm usr} \simeq \alpha^{-1} \epsilon_1^{\rm sr}\quad \Rightarrow \quad \mathcal{P}_{\mathcal{R}} \geq \alpha \, \mathcal{P}^{\rm sr}_{\mathcal{R}},$$

\noindent This condition is both necessary and sufficient for enhancement. Importantly, while $\epsilon_{2}$ reflects  the logarithmic change of the first slow-roll parameter, an attenuation of the $\epsilon_1$ necessitates $\epsilon_{2}$ takes on negative values. Therefore, while the conventional USR regime with $\epsilon^{\rm usr}_{2} \lesssim -6$ naturally leads to amplification, our analysis suggests that even a milder condition: $0 > \epsilon^{\rm usr}_{2} > -6$ can also yield the attenuation of $\epsilon_1$ and therefore the amplification of the scalar spectra. More generally, {\bf we find that any negative value of $\epsilon^{\rm usr}_{2},$ irrespective of whether it remains constant, leads to amplification. This broadens the range of viable enhancement mechanisms beyond conventional USR.} Consequently, such amplification may contribute to the production of PBHs and PGWs. This is the central finding of our work.

To demonstrate the above findings, in this article, we assume for simplicity that $\epsilon^{\rm usr}_{2}$ remains constant throughout the USR phase. Additionally, we also consider the spectral tilt to be flat and de-Sitter like in the first SR region, while it is slightly tilted in the second SR region. This allows for a more straightforward {\bf analytical yet realistic} treatment of the spectrum, avoiding the complexities of numerical analysis. {\bf We explore a range of $\epsilon_{2}^{\rm usr}$ values from $-1$ to $-10$ to examine the resulting amplification. Our findings reveal that, for $0 > \epsilon^{\rm usr}_{2} > -3,$ the enhancement is confined to sub-horizon scales, whereas for $\epsilon^{\rm usr}_{2} \leq -3,$ both sub-horizon and super-horizon amplification occurs.} In all cases, we analyze the resultant enhancement and conclude that a sufficient boost in the power spectrum is achievable, with significant implication for the production of PGWs and PBHs.

The article is structured as follows. In the next section \ref{sec:gen_eqn}, we introduce fundamental background equations, along with the equations governing curvature perturbation and define the scalar power spectrum necessary for our analysis. In Sec. \ref{sec:conventional-ampl-curv}, we review the conventional slow-roll paradigm, the freezing of curvature perturbations, and the traditional description of ultra-slow-roll (USR), where perturbations grow on super-horizon scales. In Sec. \ref{sec:proposed-amp-curv}, we present our main result, demonstrating that sub-horizon amplification alone can enhance the spectrum. Sec. \ref{sec:3-phase-sol} further explores this mechanism by analyzing different negative values of $\epsilon^{\rm USR}_{2}$ within a simplified setup. In Sec. \ref{sec:PGW} and Sec. \ref{sec:PBH}, we investigate the implications of such amplification for the production of PGWs and PBHs, comparing different values of $\epsilon_2.$ Finally, we summarize our result and provide concluding remarks in the Sec. \ref{sec:conclu}.

We begin by outlining some conventions and symbols used throughout this work. We adopt natural units, setting $\hbar = c = k_{B} = 1$,and define the reduced Planck mass as $M_{\rm pl} \equiv (8 \pi G)^{-1/2} = 1.$ The spacetime metric follows the signature $(-,+,+,+).$ Greek indices $(\mu,\nu,\cdots)$ are contracted using the metric tensor $g_{\mu \nu},$ while Latin indices $(i,j,\cdots)$ are contracted using the Kronecker delta $\delta_{i j}.$ We denote partial and covariant derivatives by $\partial$ and $\nabla,$ respectively. Time derivatives with respect to cosmic time are represented by overdots (e.g., $\dot{\phi}$),while those with respect to conformal time are denoted by primes (e.g., $\phi^\prime$). These definitions are used in the context of a Friedmann-Lema\^{\i}tre-Robertson-Walker (FLRW) background metric.

\section{General Equations} \label{sec:gen_eqn}
Before going into the details, let us first discuss the model we consider and the corresponding equations for the background as well as for the perturbations. In this article, for simplicity, we consider the simplest model of inflation where a single canonical scalar field $\phi$ is minimally coupled to the gravity with a potential $V(\phi)$:
\begin{eqnarray}
\label{eq:action-minimal}
    S=\frac{1}{2}\int {\rm d}^4 x \sqrt{-g} \left(R - g^{\mu \nu} \partial_{\mu} \phi \partial_{\nu} \phi -2 V(\phi)\right),
\end{eqnarray}
where $R$ is the Ricci scalar. The field equations corresponding to the metric as well as the scalar field can be obtained as:
\begin{eqnarray}
\label{eq:Eins-eq}
   R_{\mu \nu} - \frac{1}{2} g_{\mu \nu} R &=& T_{\mu \nu (\phi)},
    \\ 
    \label{eq:cont-eq}
    \nabla_{\mu} T^{\mu \nu}_{(\phi)} &=& 0.
\end{eqnarray}
Here, $T_{\mu \nu (\phi)}$ is the stress-energy tensor corresponding to the $\phi$ field which  is defined as:
\begin{eqnarray}
    T_{\mu \nu (\phi)} = \partial_{\mu}\phi \partial_{\nu}\phi - g_{\mu \nu}\left(\frac{1}{2}\partial_{\lambda}\phi \partial^{\lambda}\phi + V(\phi)\right).
\end{eqnarray}
Using the FLRW line element, which describes the homogeneous and isotropic Universe in cosmic time $t$ is defined as:
\begin{eqnarray}
    {\rm d}s^2 = -{\rm d}t^2 +a^2(t) {\rm d \bf x}^2
\end{eqnarray}
where, $a(t)$ is the scalar factor, the above Eqs. \eqref{eq:Eins-eq}, \eqref{eq:cont-eq} are reduced to:
\begin{eqnarray}
\label{eq:back-energy}
   && 3 H^2 = \frac{1}{2}\dot{\phi}^2 + V(\phi),
    \\
    \label{eq:back-hdot}
    &&\dot{H} = -\frac{1}{2}\dot{\phi}^2,
    \\
    \label{eq:back-phi}
    &&\ddot{\phi} + 3H\dot{\phi} + V,_{\phi} = 0,
\end{eqnarray}
where, $H = \dot{a}/a$ is the Hubble parameter and $A,_{x} \equiv \partial A/ \partial x$ , $\dot{A} \equiv \partial A/ \partial t$.

The evolution equation of the (scalar) curvature perturbation at the linear, on the other hand, can be obtained in conformal time $\eta \equiv \int {\rm d}t/a(t)$ as:
\begin{eqnarray}
\label{eq:curv-eq}
    {\mathcal{R}}''_{k} + 2 \frac{z'}{z} {\mathcal{R}}'_{k}+ {k^2}{\mathcal{R}}_{k} = 0,
\end{eqnarray}
which, by using $u_{\bf k} \equiv z {\mathcal{R}_{\bf k}},$ can be re-written in terms of the Mukhanov-Sasaki variable (scalar vacuum fluctuations) as:
\begin{eqnarray}
    \label{eq:mukhanov-sasaki}
    u''_{k} + \left(k^2 - \frac{z''}{z}\right)u_{k} = 0,\quad   z \equiv a \frac{\dot{\phi}}{H} = a\sqrt{2 \epsilon_1},
\end{eqnarray}
where, $\epsilon_1 \equiv -\dot{H}/H^2$ is the first slow-roll parameter. With the help of the background Eqs. \eqref{eq:back-energy}, \eqref{eq:back-hdot} and \eqref{eq:back-phi}, one can solve the above Mukhanov-Sasaki equation, and therefore the evolution of the curvature perturbation, \viz Eq. \eqref{eq:curv-eq}. Then the power spectrum for the curvature perturbation can be obtained as
\begin{eqnarray}
\label{eq: power-eq}
    \mathcal{P}_{{\mathcal{R}}}(k) = \frac{k^3}{2 \pi^2} |{\mathcal{R}}_{k}(\eta_{\rm e})|^2.
\end{eqnarray}
Here $\eta_{\rm e}$ denotes the end of inflation. Before discussing further specific inflationary solution in the next section, one must understand that, solving the Mukhanov-Sasaki equation is essential as it allows us to impose the initial conditions, and in this article, we confine ourselves with only the Bunch-Davies conditions, which we shall discuss in the next section.

\section{Conventional method of amplification of the curvature perturbation}\label{sec:conventional-ampl-curv}

Since the equation of the curvature perturbation as well as the Mukhanov-Sasaki equation depends on $z \equiv a \sqrt{2 \epsilon_1},$ it is crucial to understand the background dynamics governed by the Eqs. \eqref{eq:back-energy}, \eqref{eq:back-hdot} and \eqref{eq:back-phi}. These equations can be characterized by the slow-roll parameters, defined as:
\begin{eqnarray}\label{eq:def-slow-roll}
	\epsilon_1 \equiv- \frac{\dot{H}}{H^2} = \frac{\dot{\phi}^2}{2 H^2}, \quad \epsilon_2 \equiv \frac{\dot{\epsilon_1}}{H \epsilon_1} = 2 \left(\frac{\ddot{\phi}}{H \dot{\phi}} - \frac{\dot{H}}{H^2}\right).
\end{eqnarray}
Slow-roll inflation occurs when the amplitudes above two slow-roll parameters becomes sufficiently small, i.e., $|\epsilon_1|, |\epsilon_2| \ll 1,$ implying that the acceleration and velocity of the field are extremely small, i.e., $\ddot{\phi} \ll H \dot{\phi}$ and $\dot{\phi}^2 \ll H^2. $ As a result, the Eqs. \eqref{eq:back-energy} and \eqref{eq:back-phi} can be approximated as:

\begin{eqnarray}\label{eq:slow-roll-eqs}
	3  H^2 \simeq V, \quad 3 H \dot{\phi} \simeq V_{\phi}.
\end{eqnarray}
These conditions arise when the scalar field potential satisfies the condition:
\begin{eqnarray}
	\frac{V_{\phi, \phi}}{V},~ \frac{V_{\phi}}{V} \ll 1.
\end{eqnarray}
These two equations \eqref{eq:slow-roll-eqs} are known as the slow-roll equations. In this case, $\epsilon_1 \ll 1$ ensures Hubble parameter to remain nearly constant, leading to the near-exponential expansion of the universe. In this scenario, one can systematically solve the background equations as well as the evolution of the curvature perturbation, and the amplitude of the power spectrum can be obtained as

\begin{eqnarray}\label{eq:slow-roll-power-spectrum}
	\mathcal{P}_{\mathcal{R}}  \simeq \frac{H^2}{8 \pi^2 \epsilon_1}.
\end{eqnarray}
Since $\epsilon_1, \epsilon_{2}$ and $H$ remains nearly constant during the slow-roll, and therefore, the amplitude remains approximately the same. Observations constrain its value at the pivot scale $k = 0.05$ Mpc$^{-1}$  as $\mathcal{P}_{\mathcal{R}} \simeq 2.1 \times 10^{-9}.$

However, to generate  significant secondary gravitational waves as well as primordial black holes, scalar power spectrum must be amplified to the order $10^{-3} - 10^{-2}.$ Clearly, slow-roll conditions cannot achieve this, necessitating the violation of slow-roll conditions. 

One simple yet efficient way, and as a result, a widely used approach to achieve it is to introduce a new condition where the potential is nearly flat, or around a inflection point where:

\begin{eqnarray}
	V_{\phi} \simeq 0.
\end{eqnarray} 
As a consequence, Eq. \eqref{eq:back-phi} are forced to deviate from the slow-roll form to the newly modified from:
\begin{eqnarray}
	\ddot{\phi} + 3 H \dot{\phi} \simeq 0,
\end{eqnarray}
which leads to the background evolution:
\begin{eqnarray}
	\dot{\phi} \propto a^{-3},\quad \epsilon_1 \propto a^{-6}, \quad \epsilon_2  \simeq -6.
\end{eqnarray}
Contrary to the slow-roll condition, in this case, since $\epsilon_1$ decreases rapidly, leading to the rapid slow-down in the field velocity. This regime is known as the {\bf ultra slow-roll inflation}. In this case, $$\epsilon_1 \ll 1, \quad \epsilon_{2} \simeq - 6.$$ While the Hubble parameter remains nearly constant, the fast attenuation of $\epsilon_1$ leads to an increase in the scalar power spectrum, as inferred from Eq. \eqref{eq:slow-roll-power-spectrum}.

To further analyze this, we consider the super-horizon solution of the curvature perturbation equation, i.e., Eq. \eqref{eq:curv-eq}:
\begin{eqnarray}
    \mathcal{R}_{\rm sup}^{\prime \prime} + 2 \frac{z^\prime}{z}\mathcal{R}_{\rm sup}^\prime \simeq 0,
\end{eqnarray}
which, if $a(\eta) = -1/(H \eta),\ \epsilon_1 \propto a^{\epsilon_{2}} \propto (-\eta)^{-\epsilon_{2}},$ becomes:
\begin{eqnarray}
	\mathcal{R}_{\rm sup}^{\prime \prime} - \frac{2 + \epsilon_2}{\eta} \mathcal{R}_{\rm sup}^\prime \simeq 0.
\end{eqnarray}
and the solution takes the form:
\begin{eqnarray}\label{eq:curv-sup-eq}
	\mathcal{R}^{\rm sr}_{\rm sup} \simeq C_1(k) + C_2(k)\,\left(\left(\frac{\eta}{\eta_\ast}\right)^{3 + \epsilon_{2}}-1\right),
\end{eqnarray}
where $C_1(k)$ and $C_2(k)$ are determined by matching the sub-horizon solution at the horizon crossing: $\eta_\ast \simeq - 1/k$. Therefore, {\bf during slow-roll, as $\epsilon_{2} \ll 1$,} curvature perturbation freezes over the super-horizon scale. However, in {\bf conventional ultra slow-roll with $\epsilon_2 \simeq -6,$} the super-horizon solution becomes
\begin{eqnarray}
	\mathcal{R}^{\rm usr}_{\rm sup}  \simeq C_1(k) + C_2(k) \left(\left(\frac{\eta}{\eta_\ast}\right)^{-3 }-1\right) \sim \left(\frac{a}{a_\ast}\right)^{3}
\end{eqnarray}
implying that the curvature perturbation grows with the expansion of the universe (see, for instance, Refs. \cite{Byrnes:2018txb,Ng:2021hll,Cheng:2018qof,Karam:2022nym,Hazra:2012yn} for detailed analysis). As a consequence, the scalar power-spectrum during the ultra slow-roll phase amplifies significantly: $$\mathcal{P}^{\rm usr}_{\mathcal{R}} \simeq  \left(\frac{a}{a_\ast}\right)^6\, \mathcal{P}^{\rm sr}_{\mathcal{R}}.$$

However, the amplification does not hold for all values of $k$ and it depends on factors such as the duration and onset of ultra slow-roll, which impacts the two integration constants $C_1(k)$ and $C_2(k).$ Furthermore, in some models, where $\epsilon_2$ is even smaller than $-6,$ an even higher amplification of the power spectrum can be achieved due to increased super-horizon growth of the curvature perturbation. {\bf This is the traditional enhancement method of curvature power spectrum.}


\section{Proposed method: including sub-horizon amplification}\label{sec:proposed-amp-curv}

As discussed before, in the conventional picture, the amplification of the curvature perturbation typically occurs during its evolution on super-horizon scales. This naturally raises the question: {\bf can the amplification also occur on sub-horizon scales?} To answer this, we must examine the behavior of the curvature perturbation in the sub-horizon scales. The Mukhanov-Sasaki equation, i.e, Eq. \eqref{eq:mukhanov-sasaki},  in this regime ($k^2 \gg z^{\prime \prime}/z$) simplifies to

\begin{eqnarray}
	u^{\prime \prime}_{\rm sub} + k^2 u_{\rm sub} \simeq 0.
\end{eqnarray}
Using the Bunch-Davies initial condition, the solution becomes
\begin{eqnarray}
	u_{\rm sub} \simeq \frac{1}{\sqrt{2 k}}\,e^{- i k \eta} \quad \Rightarrow \quad \mathcal{R}_{\rm sub}   \simeq \frac{u_{\rm sub}}{a \sqrt{2 \epsilon_1}}.
\end{eqnarray}
In the slow-roll scenario, $\epsilon_1$ remains nearly constant. As a consequence, at the sub-horizon scale, while the amplitude of the Mukhanov-Sasaki variable remains nearly constant, the curvature perturbation $\mathcal{R}_k$ decays as $$\mathcal{R}^{\rm sr}_{\rm sub} \propto a^{-1}.$$
As the perturbation mode $k$ approaches the horizon, its amplitude continues to decrease inversely with the scale factor. Once the mode exits the horizon, the curvature perturbation effectively freezes, as  discussed in the previous section. Therefore, the amplitude of the curvature perturbation for any given mode $k$ at the end of inflation approximately matches its value at the horizon-crossing:

\begin{eqnarray}
	\left.\mathcal{R}_{\rm sup}^{\rm sr} \simeq \mathcal{R}^{\rm sr}_{\rm sub}\right|_{\ast} = \frac{1}{2 a_\ast \,k^{1/2}\sqrt{\epsilon_1} } = \frac{H}{2 \,k^{3/2}\sqrt{\epsilon_1}}
\end{eqnarray}
where $\ast$ denotes horizon crossing, i.e., $k \simeq a_\ast H.$ Accordingly, the power-spectrum thus can be obtained as

\begin{eqnarray}
\mathcal{P}_{\mathcal{R}} \simeq \frac{k^3}{2 \pi^2} \left|\mathcal{R}^{\rm sr}_{\rm sup}\right|^2 \simeq \frac{H^2}{8 \pi^2 \epsilon_1},
\end{eqnarray}
which is consistent with the standard slow-roll result, as seen in Eq. \eqref{eq:slow-roll-power-spectrum}.

However, moving beyond the slow-roll approximation by allowing $\epsilon_1$ to vary alters the evolution of curvature perturbations even before horizon crossing. Suppose $\epsilon_1$ evolves as $\epsilon_1 \propto \left(a/a_0\right)^{-\alpha} $, where $\alpha$ is a positive constant. In this case, as the amplitude of $u_k$ in the sub-horizon limit remains nearly constant, $\mathcal{R}_k$ varies as 
\begin{eqnarray}
	\mathcal{R}_{\rm sub} \propto z^{-1} = \left(a \sqrt{2 \epsilon_1}\right)^{-1} =  a^{(\alpha/2 -1)} \simeq  a^{\alpha/2}\,\mathcal{R}_{\rm sub}^{\rm sr} 
\end{eqnarray}
This leads to an amplification of the power spectrum purely due to sub-horizon evolution:
\begin{eqnarray}
	\mathcal{P}_{\mathcal{R}} \sim a^\alpha\ \mathcal{P}^{\rm sr}_{\mathcal{R}} .
\end{eqnarray}
From the Planck observation \cite{Planck:2018jri}, the slow-roll power spectrum is approximately $\mathcal{P}^{\rm sr}_{\mathcal{R}} \sim 2.1 \times 10^{-9}$\footnote{This value corresponds to the pivot scale $k_\ast = 0.05$ Mpc$^{-1}.$ Under standard slow-roll inflation, both $H$ and $\epsilon_1$ remain nearly constant throughout, so for simplicity, we assume a similar order of magnitude for other modes.}. Thus, to achieve a power spectrum of order $\sim 10^{-2},$ we estimate the required number of e-fold $\Delta N$ for ultra slow-roll regime as

$$\alpha \Delta N \simeq \log \left(\frac{\mathcal{P}_{\mathcal{R}}}{\mathcal{P}_{\mathcal{R}}^{\rm sr}}\right) \simeq 15.38.$$
As is now obvious, in this case, the second slow-roll parameter becomes $\epsilon_2  = - \alpha,$ and it is evident that a large enhancement of the power spectrum can be achieved without requiring the conventional ultra slow-roll condition $\epsilon_2 = -6$. In fact, even a mild deviation such as $\epsilon_2 = -1$ may suffice.

Moreover, even in more general scenario where $\epsilon_{2}$ evolves dynamically --- as is common in various potential models used in ultra slow-roll inflation --- the key point is that the sub-horizon power spectrum exhibits an inverse dependence on the first slow-roll parameter, i.e.,  $\mathcal{P}_{\mathcal{R}} \propto \epsilon_1^{-1}.$ Therefore, we can express this amplification in terms of $\epsilon_1$ as:
\begin{eqnarray}
\epsilon_1 \simeq \epsilon_1^{\rm sr} \left(\frac{\mathcal{P}^{\rm sr}_{\mathcal{R}}}{\mathcal{P}_{\mathcal{R}}}\right) \simeq 10^{-7} \times \epsilon_1^{\rm sr} 
\end{eqnarray}
Additionally, on super-horizon scales, the condition $\epsilon_2 \leq -3$ is sufficient to induce further growth (see, for instance, Eq. \eqref{eq:curv-sup-eq}), as opposed to the conventional ultra slow-roll criteria of $\epsilon_{2} \lesssim -6.$ The above result simply demonstrate that, {\bf irrespective of the super-horizon evolution, significant amplification of the power spectrum can be realized purely through sub-horizon evolution, provided that $\epsilon_2$ is negative and of order unity, allowing $\epsilon_1$ to reach values as small as $10^{-7} \times \epsilon_1^{\rm sr}$.}

{\bf Thus, the ultra slow-roll --- traditionally understood as a phase where the scalar field rolls extremely slowly (i.e., $|\phi_N| \ll 1$) --- can be more generally reinterpreted as an epoch during which the second slow-roll parameter $\epsilon_2$ is negative for a brief period, thereby enabling $\epsilon_1$ to decrease sufficiently to enhance the curvature power spectrum to observable levels. This mechanism, in turn, may lead to the production of significant PGWs as well as PBHs, as we will explore in the subsequent section.} 

To gain a clearer understanding of the underlying dynamics and the resulting form of the scalar power spectrum, we will next examine a simplified scenario in which $\epsilon_{2}$ is held constant and negative over a brief period. This will allow us to analyze, in detail, the consequent amplification of the curvature power spectrum.


\section{Detailed analysis of the curvature perturbation in the proposed ultra-slow condition}
\label{sec:3-phase-sol}
Let us now examine the leading order solution for curvature perturbations during the early universe. In our case, we shall follow the conventional approach by dividing the inflationary period into three distinct regimes: 
\begin{itemize}
	\item[1.] {\bf Initial Slow-Roll Regime:} In this early phase, both the first and second slow-roll parameters are small, i.e., $\epsilon_1, \epsilon_2 \ll 1$. It is during this regime that the CMB pivot scale exits the horizon.
	\item[2.] {\bf Ultra Slow-Roll Phase:} This is a brief intermediate phase characterized by $\epsilon_1 \ll 1$, while $\epsilon_2$ is negative and of order unity, i.e., $\epsilon_2 \sim \mathcal{O}(1)$. Although $\epsilon_2$ may, in general, be time-dependent, we assume it to be constant for simplicity.
	\item[3.] {\bf Final Slow-Roll Regime:} After the ultra slow-roll phase, the system transitions back to the standard slow-roll regime, which continues until the end of inflation. In this phase, instead of neglecting the slow-roll parameters $\epsilon_1$ and $\epsilon_2$, we take $\epsilon_2 = 1/3$ as a representative small value --- though any value less than unity would be suitable. This choice reflects the fact that $\epsilon_2$ contributes to the spectral tilt, deviating from the exact scale invariance of a pure de Sitter universe. A small but nonzero tilt is essential for a realistic inflationary spectrum.
\end{itemize}
Throughout all three phases, since $\epsilon_1$ remains very small, the background expansion can be well approximated by a quasi-de Sitter form: $a \simeq -1/ H \eta,$  and the slow-roll parameters can be collectively expressed as:

\begin{eqnarray}\label{eq:eps-three-regimes}
	\epsilon_{2} &=& \left\{
		\begin{aligned}
			& 0, \hspace{80pt} \eta < \eta_1 \\
			& - \epsilon^{\rm usr}_{2}, \hspace{56pt} \eta_1 < \eta < \eta_2\\
			& \frac{1}{3}, \hspace{79pt} \eta > \eta_2
		\end{aligned}\right.\\
    \epsilon_1 &=& \left\{
    \begin{aligned}
        & \epsilon_{\rm sr}, \hspace{100pt} \eta \leq \eta_1\\
        & \epsilon_{\rm sr} \left(\frac{\eta}{\eta_1}\right)^{-\epsilon^{\rm usr}_2}, \hspace{47 pt} \eta_1 \leq \eta \leq \eta_2 \\
        & \epsilon_{\rm sr} \left(\frac{\eta_2}{\eta_1}\right)^{-\epsilon^{\rm usr}_2} \left(\frac{\eta_2}{\eta}\right)^{1/3}. \hspace{8 pt} \eta \geq \eta_2
    \end{aligned}\right.,
\end{eqnarray}
where, the duration of the ultra slow-roll epoch is $\Delta N = \log \left(\frac{a(\eta_2)}{a(\eta_1)}\right) = \log \left(\frac{\eta_1}{\eta_2}\right).$ With this in mind, let us now analyze the solution of the curvature perturbations across these three regimes.

\subsection{First regime: slow-roll (SR-I)}\label{sec:slow-roll-reg-I}
In the initial regime, corresponding to the standard slow-roll phase, we apply the leading-order approximation by neglecting both $\epsilon_1$ and $\epsilon_2$ (i.e., assuming $\epsilon_1, \epsilon_2 \ll 1$). Under the leading-order slow-roll approximation, the Mukhanov-Sasaki equation \eqref{eq:mukhanov-sasaki} simplifies to

\begin{eqnarray}\label{eq:MS_SR1}
	 u^{(1) \prime \prime}_k + \left(k^2 - \frac{2}{\eta^2}\right) u_k^{(1)} = 0,
\end{eqnarray}
where we have used $z''/z \simeq 2/\eta^{2}$. The general solution to this equation is given by

\begin{eqnarray} 
	u_k^{(1)}(\eta) \simeq \sqrt{-\eta} \left(\alpha_{1} H^{(1)}_{3/2}(-k \eta) + \beta_{1} H^{(2)}_{3/2}(-k \eta)\right), 
	\end{eqnarray}
where $H^{(1)}_\nu(x)$ and $H^{(2)}_\nu(x)$ denote Hankel functions of the first and second kind, and $\alpha_1(k), \beta_1(k)$ are the Bogoliubov coefficients determined by the choice of the initial vacuum state. Imposing the Bunch-Davies initial condition that selects the positive frequency mode in the far past, yielding $ \alpha_{1} = \frac{\sqrt{\pi}}{2}, \quad \beta_{1} = 0.$ This leads to the explicit solution

\begin{eqnarray}
	u^{(1)}_k(\eta) \simeq \frac{1}{\sqrt{2k}} \left(\frac{i}{k \eta} - 1\right) e^{-i k \eta}, \quad \eta \leq \eta_1,
\end{eqnarray}
up to an irrelevant phase factor. Consequently, the corresponding solution for the curvature perturbation $\mathcal{R}_k^{(1)} \equiv u^{(1)}_k / z$ becomes

\begin{eqnarray} \mathcal{R}^{(1)}_{k}(\eta) \simeq \frac{H}{2 k^{3/2} \sqrt{\epsilon_{\rm sr}}} (k \eta - 1) e^{-i k \eta}, \quad \eta \leq \eta_1. \end{eqnarray}
Here, we consider $\epsilon_{\rm sr} \sim 10^{-3}$ (again, as a representative small number, as is expected from small field inflationary models).

\subsection{Second regime: ultra slow-roll (USR)}\label{sec:usr-ref-II}
In the intermediate ultra slow-roll phase, $\epsilon_1 \ll 1$ and $\epsilon_{2} \sim \mathcal{O}(1).$ Therefore, in the Mukhanov-Sasaki equation, unlike the slow-roll scenario, one cannot ignore the contribution from the second slow-roll parameter as it takes the form:

\begin{eqnarray} {u''}^{(2)}_k + \left(k^2 - \frac{(2+\epsilon^{\rm usr}_{2}) (4+\epsilon^{\rm usr}_{2})}{4\eta^2}\right) u_k^{(2)} = 0. \end{eqnarray}
The general solution to this equation is

\begin{eqnarray} u^{(2)}_k(\eta) \simeq \sqrt{-\eta} \left(\alpha_{2} H^{(1)}_{\nu}(-k \eta) + \beta_{2} H^{(2)}_{\nu}(-k \eta)\right), \end{eqnarray}
where the index is defined as $\nu = (3 + \epsilon^{\rm usr}_2)/2$, and $\alpha_2, \beta_2$ are the Bogoliubov coefficients. The curvature perturbation is then given by

\begin{eqnarray} \mathcal{R}^{(2)}_k(\eta) \simeq \frac{H (-\eta)^{3/2}}{\sqrt{2 \epsilon_{\rm sr} \left(\frac{\eta_1}{\eta}\right)^{\epsilon_2}}} \left(\alpha_{2} H^{(1)}_{\nu}(-k \eta) + \beta_{2} H^{(2)}_{\nu}(-k \eta)\right). \end{eqnarray}
The expressions for the coefficients $\alpha_2$ and $\beta_2$ are obtained by applying continuity and smoothness conditions at $\eta = \eta_1$ with the solution of the curvature perturbation in the first slow-roll phase given in the previous section. 

\subsection{Third regime: slow-roll (SR-II)}
In the final stage, the system returns to the slow-roll regime. In this epoch, for simplicity, while we ignore $\epsilon_1,$ $\epsilon_{2}$ is set to be $1/3$ as a small value. Under this assumption, the Mukhanov-Sasaki equation becomes:

\begin{eqnarray} u^{(3)}_k(\eta) = \sqrt{-\eta} \left(\alpha_{3} H^{(1)}_{5/3}(-k \eta) + \beta_{3} H^{(2)}_{5/3}(-k \eta)\right). \end{eqnarray}
Accordingly, the curvature perturbation evolves as

\begin{eqnarray} \mathcal{R}^{(3)}_k(\eta) = \frac{H (-\eta)^{3/2}}{\sqrt{2 \epsilon_{\rm sr} \left(\frac{\eta_1}{\eta_2}\right)^{\epsilon_2} \left(\frac{\eta_2}{\eta}\right)^{1/3}}} \left(\alpha_{3} H^{(1)}_{5/3}(-k \eta) + \beta_{3} H^{(2)}_{5/3}(-k \eta)\right).
\end{eqnarray}
In a similar way, the Bogoliubov coefficients $\alpha_3$ and $\beta_3$ are determined by matching it with the curvature perturbation obtained the USR phase at the transition point $\eta = \eta_2$.

\subsection{Evolution and power spectrum of the curvature perturbation}

\begin{figure}[h]
	\centering
	\includegraphics[width=.7\linewidth]{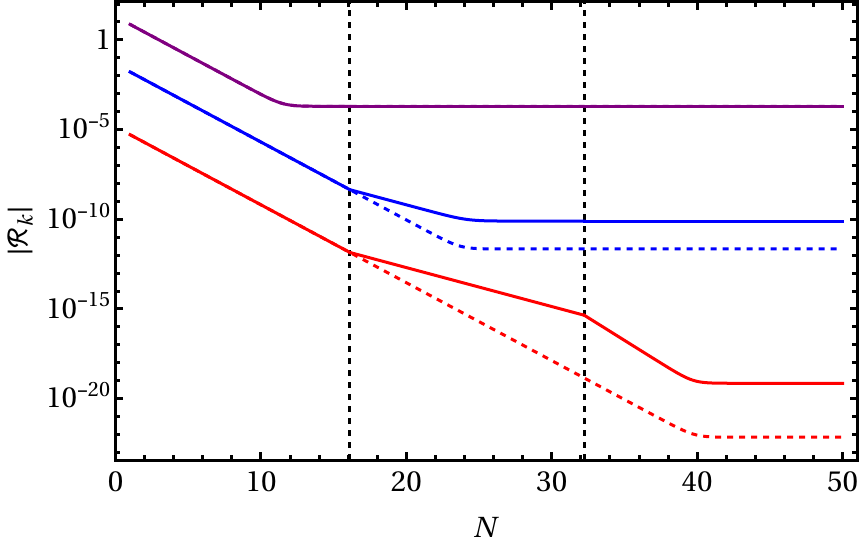}
	\includegraphics[width=.7\linewidth]{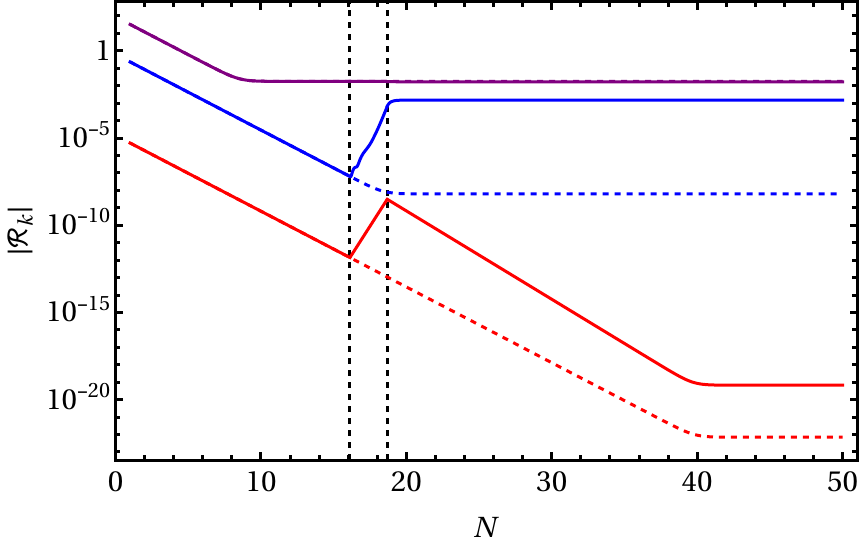}
	\caption{These plots illustrate the evolution of perturbations both in the presence (solid lines) and absence (dotted lines) of the USR phase. The \textbf{top panel} corresponds to the case with $\epsilon_2 = -1$, where the USR phase occurs between $\eta_1 = -10^{-2}\ \text{Mpc}$ and $\eta_2 = -10^{-9}\ \text{Mpc}$. The {\color{purple}purple} line shows the evolution for the mode $k = 10^{-1}\ \text{Mpc}^{-1}$, the {\color{blue}blue} line for $k = 700\ \text{Mpc}^{-1}$, and the {\color{red}red} line for $k = 10^{12}\ \text{Mpc}^{-1}$. The \textbf{bottom panel} shows the case with $\epsilon_2 = -6$, and the USR phase spans from $\eta_1 = -10^{-2}\ \text{Mpc}$ to $\eta_2 = -7.7 \times 10^{-4}\ \text{Mpc}$. Here, the {\color{purple}purple} curve corresponds to $k = 5 \times 10^{-2}\ \text{Mpc}^{-1}$, the {\color{blue}blue} to $k = 10^3\ \text{Mpc}^{-1}$, and the {\color{red}red} to $k = 2 \times 10^{12}\ \text{Mpc}^{-1}$.
	}
	\label{fig:evcurv}
\end{figure}

We now examine the evolution of the curvature perturbation in two representative SR-USR-SR scenario, characterized by different values of the second slow-roll parameter, $\epsilon^{\rm usr}_2 = -1$ and $-6$. These cases are central to our analysis, as they illustrate the sub-horizon as well as combined sub and super-horizon growth of curvature perturbations, respectively. The evolution for each case is shown in \ref{fig:evcurv}.

The curvature perturbation evolves in three distinct stages as:
\begin{eqnarray}
	\mathcal{R}_{k}(\eta) = \left\{\begin{aligned}
		&\mathcal{R}_k^{(1)} (\eta),\quad \eta < \eta_1 \\
		&\mathcal{R}_k^{(2)} (\eta),\quad \eta_1 \leq \eta \leq \eta_1 \\
		&\mathcal{R}_k^{(3)} (\eta).\quad \eta > \eta_2
	\end{aligned}\right.
\end{eqnarray}
as discussed in the previous section.

{\bf In the top panel of Fig. \ref{fig:evcurv}, we consider $\epsilon^{\rm usr}_2 = -1$,} with the USR phase spanning from $\eta_1 = -10^{-2}$ Mpc to $\eta_2 = -10^{-9}$ Mpc. The figure displays the evolution of three mode solutions corresponding to $k = 10^{-1}$ Mpc$^{-1}$ (purple), $k = 700$ Mpc$^{-1}$ (blue), and $k = 10^{12}$ Mpc$^{-1}$ (red). As discussed earlier, when $\epsilon^{\rm usr}_2 = -1$, there is no super-horizon growth of the curvature perturbation. For the mode $k = 10^{-1}$ Mpc$^{-1}$ (purple), horizon crossing occurs at $\eta_\ast = -10$ Mpc, which is well before the onset of the USR phase ($\eta_\ast < \eta_1$). As a consequence, the mode decays as $a^{-1}$ in the sub-horizon regime and subsequently freezes after horizon crossing, without experiencing any growth from the USR dynamics.

In contrast, the mode $k = 10^{12}$ Mpc$^{-1}$ (red) crosses the horizon after the end of the USR phase ($\eta_\ast > \eta_2$), and thus undergoes a relative sub-horizon amplification of $a^{1/2}$ compared to the standard slow-roll case (dotted line). The intermediate mode $k = 700$ Mpc$^{-1}$ (blue), with horizon crossing between $\eta_1$ and $\eta_2$, lies partially within the USR phase and therefore experiences moderate enhancement.

{\bf In the bottom panel, we examine the case $\epsilon^{\rm usr}_2 = -6$,} where the curvature perturbation exhibits growth even on super-horizon scales. Here, modes with early horizon crossing ($\eta_\ast \ll \eta_1$, purple) again show no amplification, as they exit the horizon before the USR phase. However, modes that cross near or within the USR interval (blue) undergo both sub-horizon and super-horizon growth. Modes with very late horizon crossing ($\eta_\ast \gg \eta_2$, red) experience only sub-horizon amplification.

Since the amplification process is now known, we can evaluate the inflationary power spectrum at the end of inflation as:

\begin{eqnarray}
	\mathcal{P}_{\mathcal{R}}  = \frac{k^3}{2 \pi^2} \left|\mathcal{R}_k^{(3)}(\eta_{\rm e})\right|^2.
\end{eqnarray}

 In the following Fig. \ref{fig:power-spectrum}, we present the resulting power spectrum for several values of $\epsilon^{\rm usr}_2 = -1, -2, -3, -6, -8$, and $-10$. As expected, for each cases, we obtain the enhanced power spectrum to our desired strength, which basically illustrates our claim. {\bf To summarize our results:}
\begin{itemize}
	\item Any negative value of $\epsilon^{\rm usr}_{2}$ leads to an enhancement of the curvature perturbation via sub-horizon amplification.
	\item Super-horizon amplification does not occur until $\epsilon^{\rm usr}_{2}$ becomes less than or equal to $-3$ (see, for instance, Eq. \eqref{eq:curv-sup-eq}).
	\item For $\epsilon^{\rm usr}_{2} \leq -3,$ both sub-horizon and super-horizon amplification mechanisms are active.
	\item The shapes of the resulting power spectra are distinct for different values of $\epsilon^{\rm usr}_{2},$ making them potentially distinguishable observationally (see Fig. \ref{fig:power-spectrum}).
	\item The more negative the value of $\epsilon^{\rm usr}_{2},$ the shorter the duration of the ultra slow-roll phase required to achieve significant power spectrum enhancement.
\end{itemize}

Thus, sub-horizon growth emerges as a key factor in enhancing curvature perturbations --- a point that is sometimes underappreciated in the literature.  {\bf This is the central result of our work.}

\begin{figure}[t]
	\centering
	\includegraphics[width=.47\linewidth]{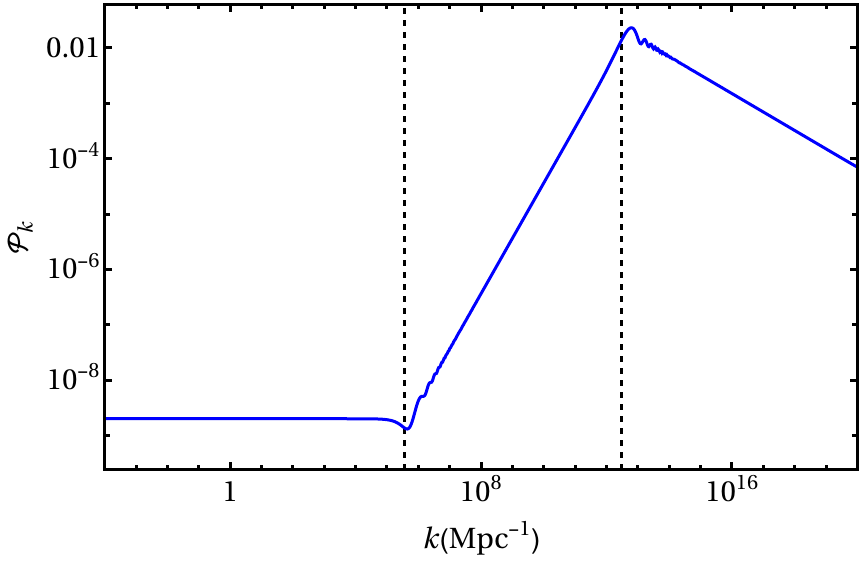}
	\includegraphics[width=.47\linewidth]{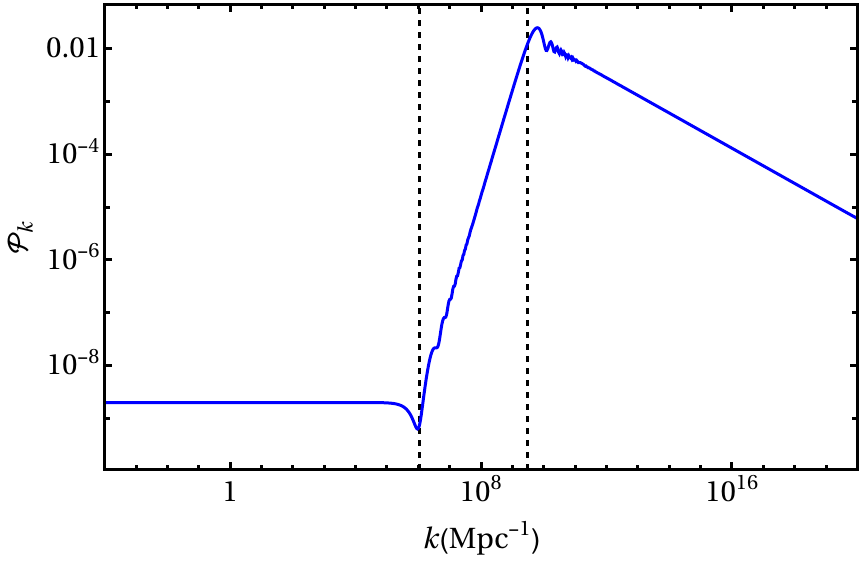}
	\includegraphics[width=.47\linewidth]{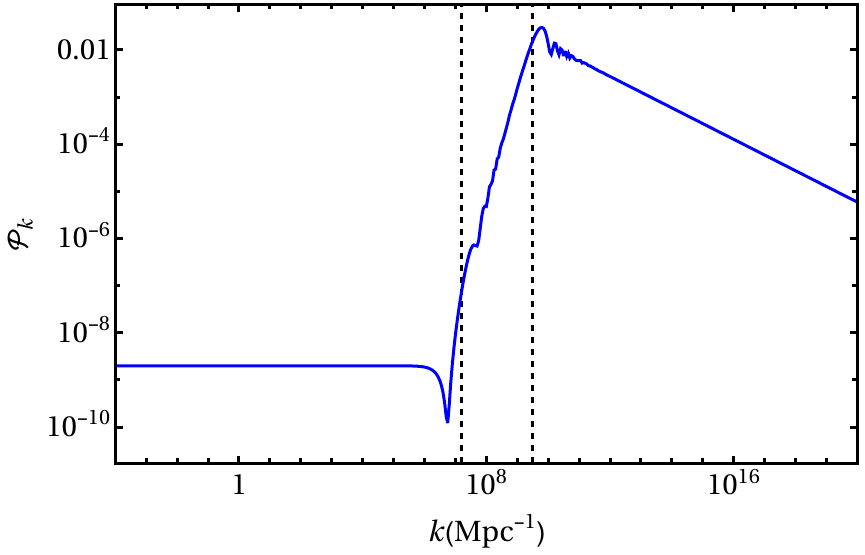}
	\includegraphics[width=.47\linewidth]{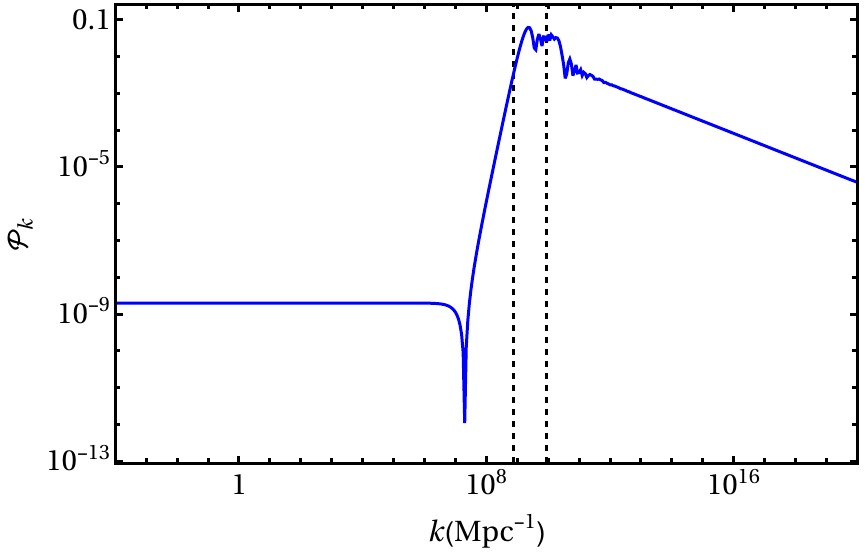}
	\includegraphics[width=.47\linewidth]{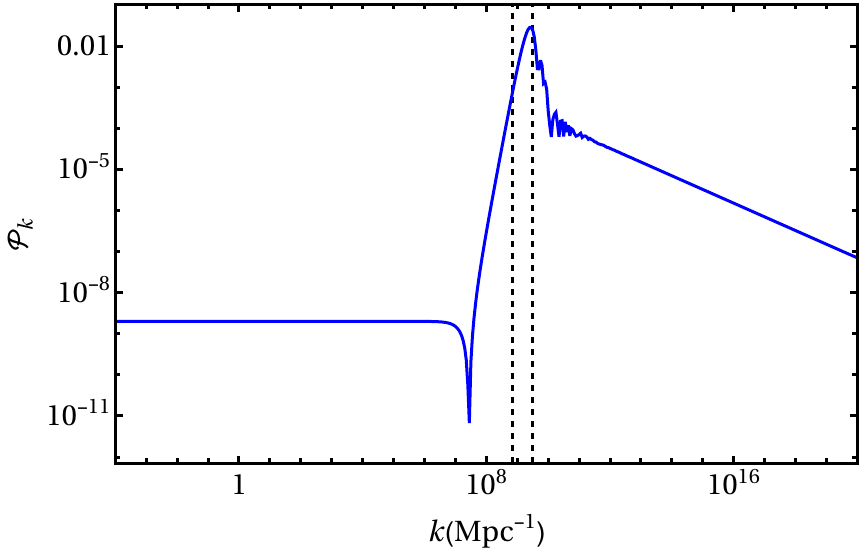}
	\includegraphics[width=.47\linewidth]{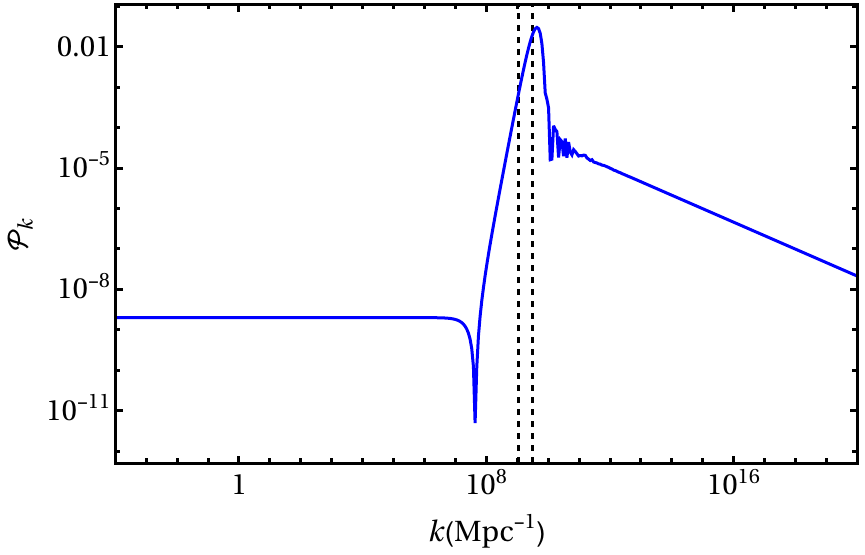}
	\caption{The power spectra are shown as functions of the wave number $k$ for different values of the second slow-roll parameter, $\epsilon^{\rm usr}_2$ during the USR phase spanning in between $\eta_1$ to $\eta_2.$ The values of $\eta_1$ and $\eta_2$ are chosen meticulously to obtain an power spectrum of order $\sim 10^{-2}$. In the \textbf{first row}, the {\bf left plot} corresponds to $\epsilon^{\rm usr}_2 = -1$ with $\eta_1 = -3 \times 10^{-6}\ \text{Mpc}$ and $\eta_2 = -3.3 \times 10^{-13}\ \text{Mpc}$, while the {\bf right plot} shows the case for $\epsilon^{\rm usr}_2 = -2$ with $\eta_1 = -9.1 \times 10^{-7}\ \text{Mpc}$ and $\eta_2 = -3.3 \times 10^{-10}\ \text{Mpc}$. In the \textbf{middle row}, the {\bf left plot} corresponds to $\epsilon_2 = -3$ with $\eta_1 = -6.5 \times 10^{-8}\ \text{Mpc}$ and $\eta_2 = -3.3 \times 10^{-10}\ \text{Mpc}$, while the {\bf right plot} illustrates the spectrum for $\epsilon_2 = -6$ with $\eta_1 = -1.25 \times 10^{-9}\ \text{Mpc}$ and $\eta_2 = -1.11 \times 10^{-10}\ \text{Mpc}$. At the \textbf{bottom row}, the {\bf left plot} corresponds to $\epsilon_2 = -8$ with $\eta_1 = -1.37 \times 10^{-9}\ \text{Mpc}$ and $\eta_2 = -3.3 \times 10^{-10}\ \text{Mpc}$, and the {\bf right plot} shows the case for $\epsilon_2 = -10$ with $\eta_1 = -9.216 \times 10^{-10}\ \text{Mpc}$ and $\eta_2 = -3.3 \times 10^{-10}\ \text{Mpc}$.}\label{fig:power-spectrum}
\end{figure}

The realization of enhanced spectra across a broad range of $\epsilon_2$ values naturally leads to the next question: {\bf what are the implications for the production of primordial gravitational waves (PGWs) and primordial black holes (PBHs)?} In the following sections, we explore these consequences in detail.

\section{Scalar-Induced Gravitational Waves from Primordial Power Spectrum} \label{sec:PGW}

Now that we have thoroughly discussed the amplification of curvature perturbations, we turn our attention to their consequences for the production of gravitational waves (GWs). Specifically, we consider the modes $k$ that re-enter the horizon during the radiation-dominated era. In this context, the relevant quantity is the energy density of the gravitational waves, denoted by $\Omega_{\rm GW}$, which is given by \cite{Kohri:2018awv, Domenech:2021ztg, Yi:2023tdk}:

\begin{equation} \label{eq:OMGW}
	\Omega_{\rm GW}(\eta , k) = \frac{1}{24} \left( \frac{k}{a H} \right)^2 \overline{\mathcal{P}_h(\eta,k)}
\end{equation}
where $\mathcal{P}_h$ is the power spectrum of tensor perturbations induced at second order, and the overline denotes the oscillation average. To compute this, we must solve the scalar-induced tensor mode equation of motion:
\begin{equation} \label{eq:SIGWeq}
	h_{k}'' + 2 \mathcal{H}h_{k}' + k^2 h_{k} = 4 S_{k}(\eta),
\end{equation}
where $\eta$ is the conformal time, the prime denotes the derivative with respect to $\eta$, $\mathcal{H} \equiv a H$ is the conformal Hubble parameter, and $S_{\mathbf{k}}(\eta)$ is the source term arising from the scalar perturbations:

\begin{equation} \label{eq:Source-term}
	S_k = \int \frac{d^3\tilde{\bf k}}{(2\pi)^{3/2}} e_{ij}(\mathbf{k}) \Tilde{k}^i \Tilde{k}^j \left[2 \Phi_{\mathbf{\tilde{k}}} \Phi_{\mathbf{k - \tilde{k}}} + \frac{1}{\mathcal{H}^2} \left(\Phi_{\mathbf{\Tilde{k}}}' + \mathcal{H} \Phi_{\mathbf{\Tilde{k}}}\right) \left(\Phi_{\mathbf{k - \Tilde{k}}}' + \mathcal{H} \Phi_{\mathbf{k - \Tilde{k}}} \right) \right].
\end{equation}
Here, $\Phi_{\mathbf{k}}$ is the Bardeen Potential and $e_{ij}(\bf k)$ is the polarization tensor. The solution of $h_{k}$ in equation \eqref{eq:SIGWeq} can be obtained using Green's function method which is the form:

\begin{equation} \label{eq:h-green-soln}
	h_{k} = \frac{4}{a(\eta)} \int^\eta G_{k}(\eta,\bar{\eta})\ a(\eta)\ S_{k}(\eta)\ d\bar{\eta}
\end{equation}
with the corresponding Green's function:	
\begin{equation} \label{eq:Green-soln}
	G_{k}(\eta,\bar{\eta}) = \frac{\sin[k(\eta-\bar{\eta})]}{k}.
\end{equation}
The power spectrum of the gravitational wave is defined through:
\begin{equation} \label{eq:power-h-induced}
	\langle h_{k} h_{k'} \rangle = \delta^3(\mathbf{k}+\mathbf{k'}) \frac{2\pi^2}{k^3} \mathcal{P}_{h}.
\end{equation}
By combining Eqs. \eqref{eq:SIGWeq}, \eqref{eq:Source-term},\eqref{eq:h-green-soln}, \eqref{eq:Green-soln} and \eqref{eq:power-h-induced}, the power spectrum of the induced GWs becomes \cite{Domenech:2021ztg,Kohri:2018awv,Yi:2023tdk,Yu:2023lmo}:
\begin{equation} \label{eq:ps-h-induced-full}
	\mathcal{P}_{h} = 4 \int^{\infty}_0 dv \int^{1+v}_{|1-v|} du \left[\frac{4v^2 - (1-u^2+v^2)^2}{4uv} \right]^2  I_{RD}^2(u,v,x) \mathcal{P}_{\mathcal{R}}{(kv)} \mathcal{P}_{\mathcal{R}}{(ku)}
\end{equation}
and accordingly, the expression for $\Omega_{GW}$ becomes:
\begin{equation} \label{eq:OMGW-induced}
	\Omega_{\rm GW} = \frac{1}{6} \left( \frac{k}{a H} \right)^2 \int^{\infty}_0 dv \int^{1+v}_{|1-v|} du \left[\frac{4v^2 - (1-u^2+v^2)^2}{4uv} \right]^2  \overline{I_{RD}^2(u,v,x\rightarrow\infty)} \mathcal{P}_{\mathcal{R}}(kv) \mathcal{P}_{\mathcal{R}}(ku).
\end{equation}
Here, $\mathcal{P}_{\mathcal{R}}$ is the primordial curvature power spectrum, $u=|\mathbf{k-\Tilde{k}}|/k$ , $v=|\mathbf{\Tilde{k}}|/k$, and $\overline{I_{RD}^2}$ is the oscillation-averaged transfer function squared. At late times $k \eta \gg 1,$ this average takes the analytical form \cite{Kohri:2018awv,Yi:2023tdk}:

\begin{multline*} \label{eq:transfer-fn}
	\overline{I_{RD}^2(u,v,x\rightarrow \infty)} = \frac{1}{2} \left[\frac{3(u^2+v^2-3)}{4u^3v^3x} \right]^2 \bigg[\left(-4uv + (u^2+v^2-3)\log\left|\frac{3-(u+v)^2}{3-(u-v)^2}\right|\right)^2 + \\\pi^2(u^2+v^2-3)^2\Theta(v+u-\sqrt{3}) \bigg]
\end{multline*}
Since the gravitational wave energy density scales like radiation, we can estimate the present-day GW energy density as:

\begin{eqnarray}
	\Omega_{\rm GW}(k, \eta_0) = \frac{\Omega_{r, 0} \Omega_{\rm GW}(k, \eta) }{\Omega_{r}(\eta)} \simeq \Omega_{r, 0} \Omega_{GW}(k, \eta),
\end{eqnarray}
where $\Omega_{r, 0}$ is the present day radiation density, and $\Omega_{r}(\eta) \simeq 1$ during the radiation-dominated epoch (the relevant modes re-enter the horizon).

As discussed in the previous section, we have demonstrated that any negative value of $\epsilon_{2}^{\rm usr}$ can significantly enhance curvature power spectrum. Using the corresponding equations, along with power spectra generated for the different values of $\epsilon_{2}^{\rm usr},$ one can compute the present day abundance of induced gravitational waves, $\Omega_{\rm GW}.$ In the following Fig. \ref{fig:GW_plot}, we plot the resulting gravitational wave energy densities for $\epsilon^{\rm usr}_{2} = -1,-2,-3,-6,-8$ and $-10$ considering various choices of $\eta_1$ and $\eta_2.$ These scenarios are then compared with the sensitivities of current and upcoming observational probes. It is evident that all the considered models yield gravitational wave signals that are within reach of these experiments, and hence can be tightly constrained or potentially detected. These results further reinforce our main conclusion: {\bf any negative value of $\epsilon_{2}^{\rm usr}$ can generate a significant gravitational wave background, which is testable with current and future observations.}
\begin{figure}[H]
	\centering
	\captionsetup{justification=justified}
	\includegraphics[width=0.47\textwidth]{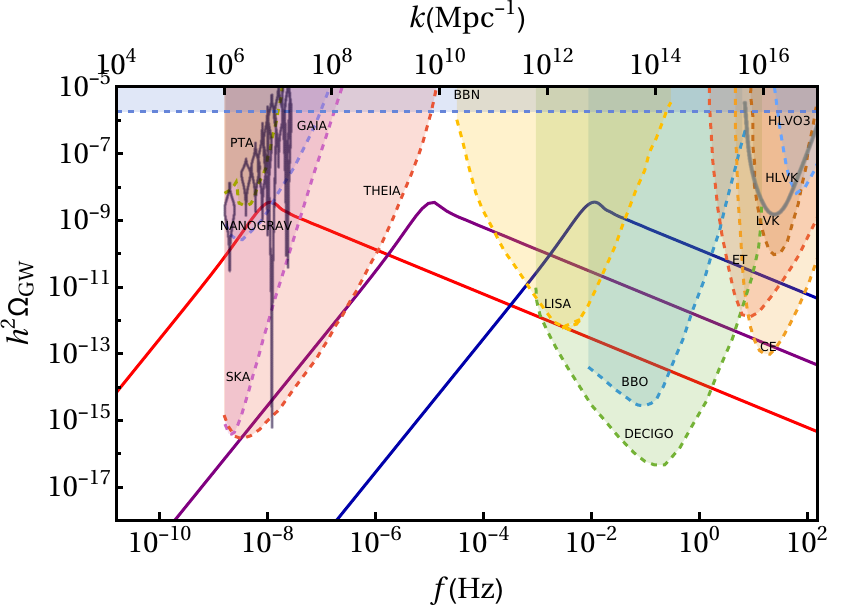} 
	\includegraphics[width=0.47\textwidth]{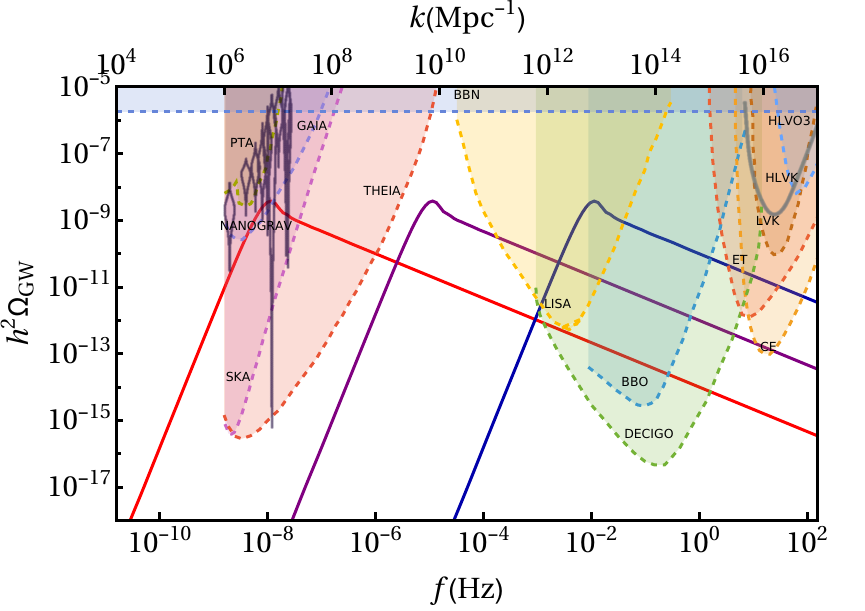} 
	\includegraphics[width=0.47\textwidth]{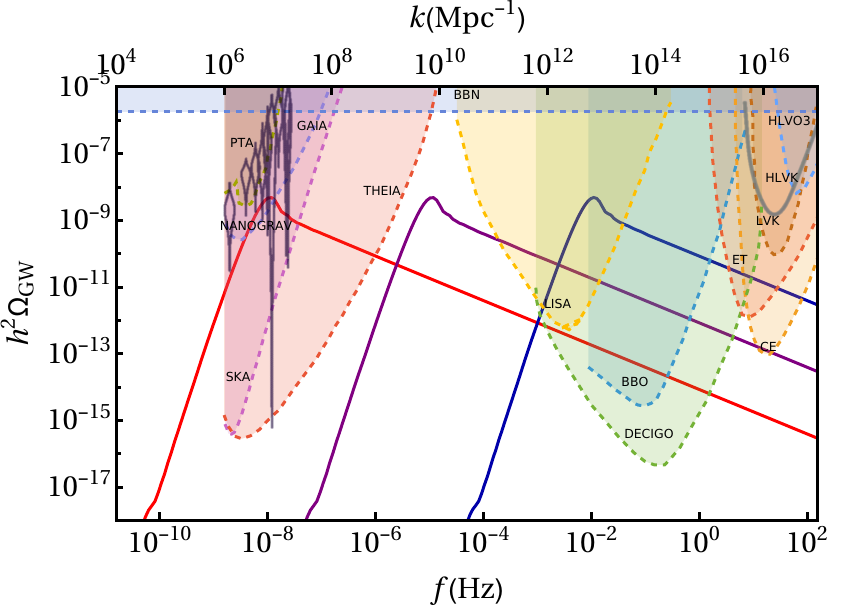} 
	\includegraphics[width=0.47\textwidth]{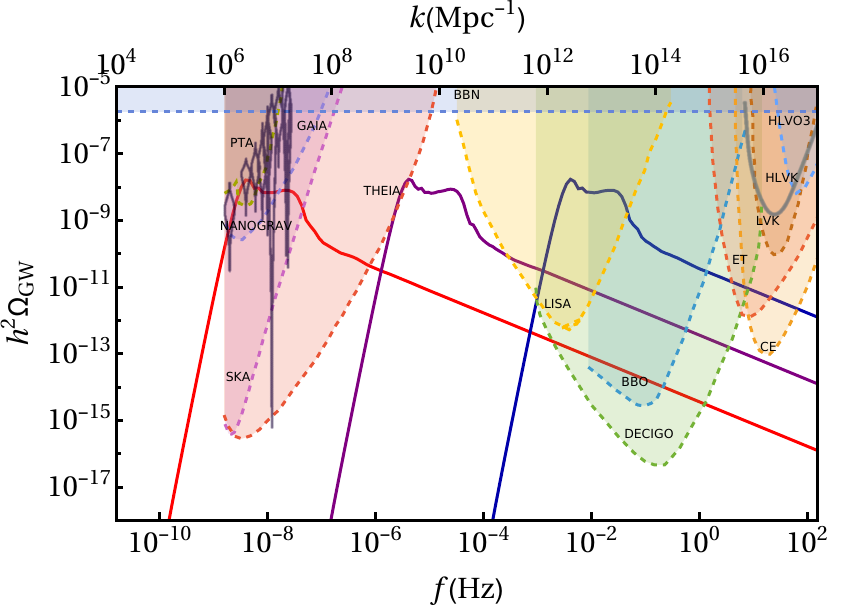} 
	\includegraphics[width=0.47\textwidth]{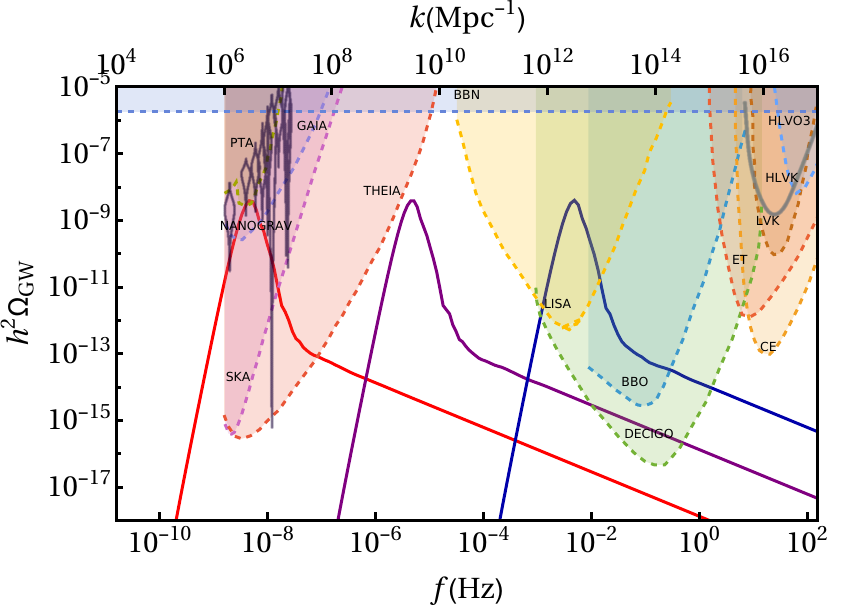} 
	\includegraphics[width=0.47\textwidth]{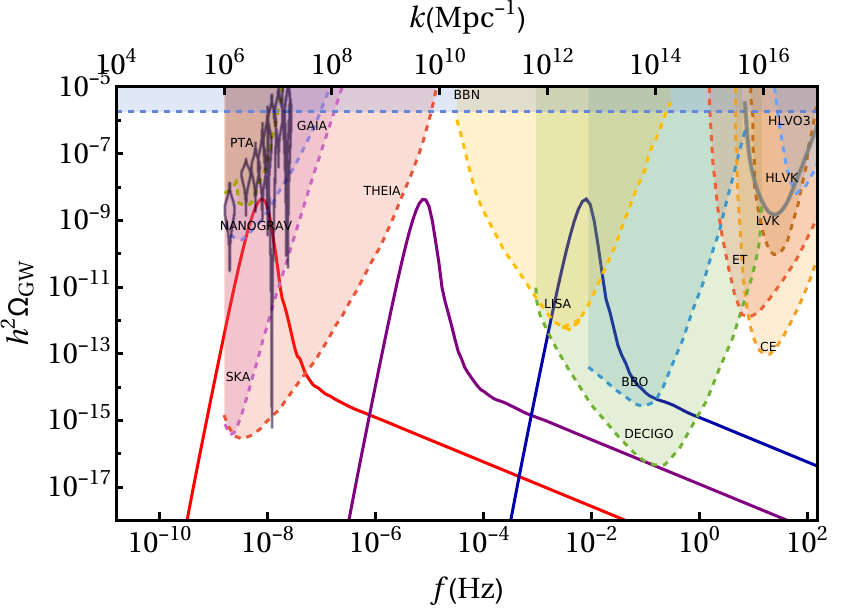} 
	\caption{The spectra of second-order induced gravitational waves ($\Omega_{\rm GW}$) are presented for various sets of $\eta_1$ (onset of the USR phase) and $\eta_2$ (end of the USR phase, fixed for spectrum of same color). The {\color{red}Red} curves correspond to $\eta_2 = -3.33 \times 10^{-7}\ \text{Mpc}$, the {\color{purple}Purple} to $\eta_2 = -3.33 \times 10^{-10} \ \text{Mpc}$, and the {\color{blue}Blue} to $\eta_2 = -3.33 \times 10^{-12} \ \text{Mpc}$.	In the \textbf{first panel}, the \textbf{left plot} shows the SIGW spectrum for $\epsilon_2 = -1$ with $\eta_1 = -2.86,\ -2.86 \times 10^{-3},\ -2.86 \times 10^{-5}\ \text{Mpc}$ ({\color{red}Red}, {\color{purple}Purple}, {\color{blue}Blue}, respectively), while the \textbf{right plot }corresponds to $\epsilon_2 = -2$ with $\eta_1 = -9.09 \times 10^{-4},\ -9.09 \times 10^{-7},\ -9.09 \times 10^{-10}\ \text{Mpc}$ ({\color{red}Red}, {\color{purple}Purple}, {\color{blue}Blue}, respectively).	In the \textbf{second panel}, the \textbf{left plot} represents the spectrum for $\epsilon_2 = -3$ with $\eta_1 = -6.33 \times 10^{-4},\ -6.33 \times 10^{-7},\ -6.33 \times 10^{-10}\ \text{Mpc}$, and the \textbf{right plot} shows results for $\epsilon_2 = -6$ with $\eta_1 = -3.77 \times 10^{-5},\ -3.77 \times 10^{-8},\ -3.77 \times 10^{-11}\ \text{Mpc}$ (colors as before).	In the \textbf{third panel}, the \textbf{left plot} displays SIGW for $\epsilon_2 = -8$ with $\eta_1 = -1.38 \times 10^{-5},\ -1.38 \times 10^{-8},\ -1.38 \times 10^{-11}\ \text{Mpc}$, and the \textbf{right plot} corresponds to $\epsilon_2 = -10$ with $\eta_1 = -9.22 \times 10^{-6},\ -9.22 \times 10^{-9},\ -9.22 \times 10^{-12}\ \text{Mpc}$ (again, {\color{red}Red}, {\color{purple}Purple}, {\color{blue}Blue}, respectively).}
	\label{fig:GW_plot}
\end{figure}

\section{PBHs Formation from inflationary Scalar Power Spectrum} \label{sec:PBH}
Primordial black holes (PBHs) can form when an over-dense region at a scale $k>10^{-2}\ \text{Mpc}^{-1}$ collapses upon re-entering the Hubble radius during the radiation-dominated era. For such collapse to occur, the curvature power spectrum must be amplified to values of $\mathcal{O}(0.01)$ is required for formation of PBHs \cite{Gangopadhyay:2021kmf}. In this section, we compute the fraction of PBHs that contribute to the dark matter today. To begin with, we consider the relation between curvature perturbation $\mathcal{R}_k$ and density contrast $\delta(t,k)$, given by \cite{Green:2004wb}:
\begin{eqnarray}
    \delta(t,k)=\frac{2(1+w)}{5+3w}\left(\frac{k}{aH}\right)^2{\mathcal{R}}_k,
\end{eqnarray}
where $w$ is the equation of state parameter. During radiation-dominated epoch ($w=1/3$), the matter power spectrum $\mathcal{P}_{\delta}(k)$ is thus related to the primordial scalar power spectrum $\mathcal{P}_{\mathcal{R}}(k)$ via:
\begin{eqnarray}
    \mathcal{P}_\delta(k)=\frac{16}{81}\ \left(\frac{k}{aH}\right)^{4}\ \mathcal{P}_{\mathcal{R}}(k).
\end{eqnarray}
Assuming a Gaussian probability distribution for the density perturbation $\delta$, which is given by
\begin{eqnarray}
    \mathbf{P}(\delta)=\frac{1}{\sqrt{2\pi\sigma^{2}(M)}}\ \exp\left(-\frac{\delta^{2}}{2\sigma^2(M)}\right),
\end{eqnarray}
where $\sigma^2(M)$ represents the variance of the density contrast, smoothed over a comoving scale $R(M) \equiv 1/k$ and is defined as:
\begin{eqnarray}
    \sigma^{2}(M) = \int_{0}^\infty d \ln(k)\ \mathcal{P}_{\delta}(k) \ \mathcal{W}^{2}(k R(M))
\end{eqnarray}
with $\mathcal{W}(k R) \equiv \exp(-k^2 R^2/2)$ being the Gaussian window function. PBHs are formed in regions where $\delta > \delta_{c}$, for a threshold value $\delta_{c}$ known as the critical density contrast. The fraction of energy density in PBHs at the time of their formation is calculated using the Press-Schechter formalism \cite{Baumann:2022mni}:
\begin{eqnarray}
    \beta(M_{\text{PBH}})=\int_{\delta_c}^{\infty}\ d\delta\ \mathbf{P}(\delta)\simeq\ \frac{1}{2}\left[\text{Erfc}\left(\frac{\delta_{c}}{\sqrt{2\ \sigma^{2}(M)}}\right)\right]\ \simeq\ \frac{\gamma}{\sqrt{2\ \pi}\nu(M)}\ \exp\left[-\frac{\nu^2(M)}{2}\right]
\end{eqnarray}
where $\nu=\delta_{c}/\sigma(M)$ is the peak threshold parameter and $\rm{Erfc}(x)$ is the complementary error function. The critical threshold $\delta_{c}$ is model-dependent and significantly impacts PBHs formation. For our analysis, we consider $\delta_{c} = 1/3$ \cite{1975ApJ...201....1C}.

\begin{figure}[H]
 \centering
    \captionsetup{justification=justified}
    \includegraphics[width=0.47\textwidth]{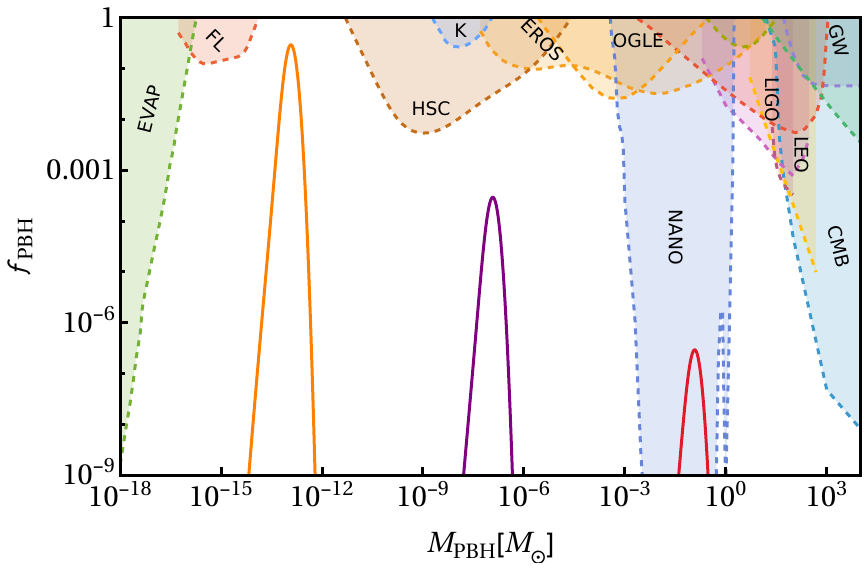} 
    \includegraphics[width=0.47\textwidth]{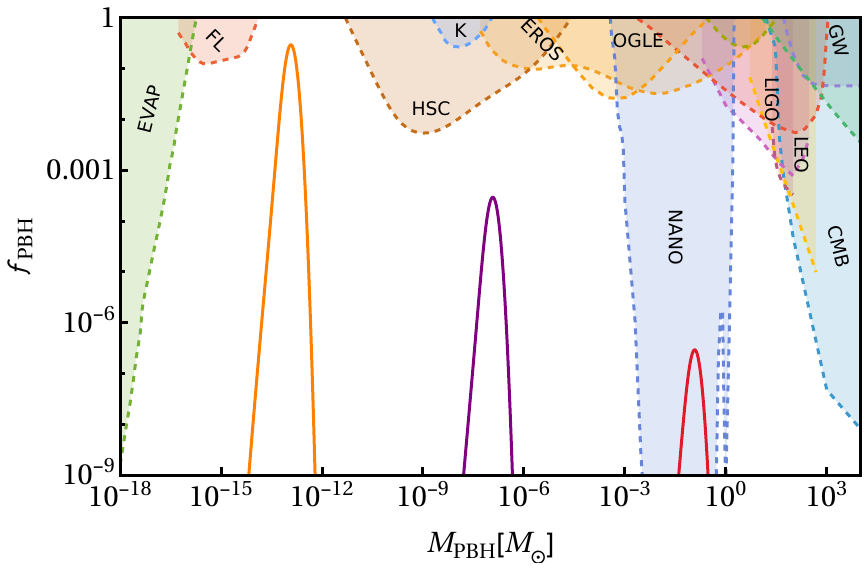} 
    \includegraphics[width=0.47\textwidth]{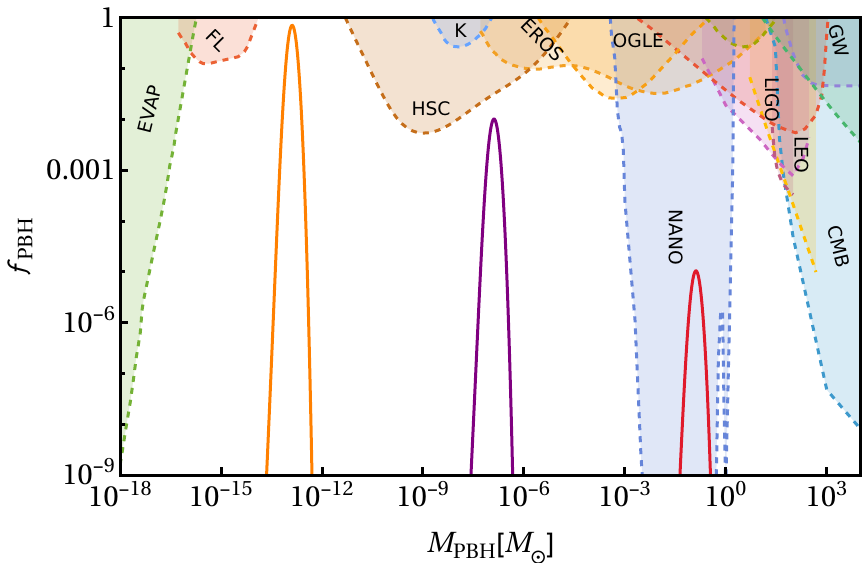} 
    \includegraphics[width=0.47\textwidth]{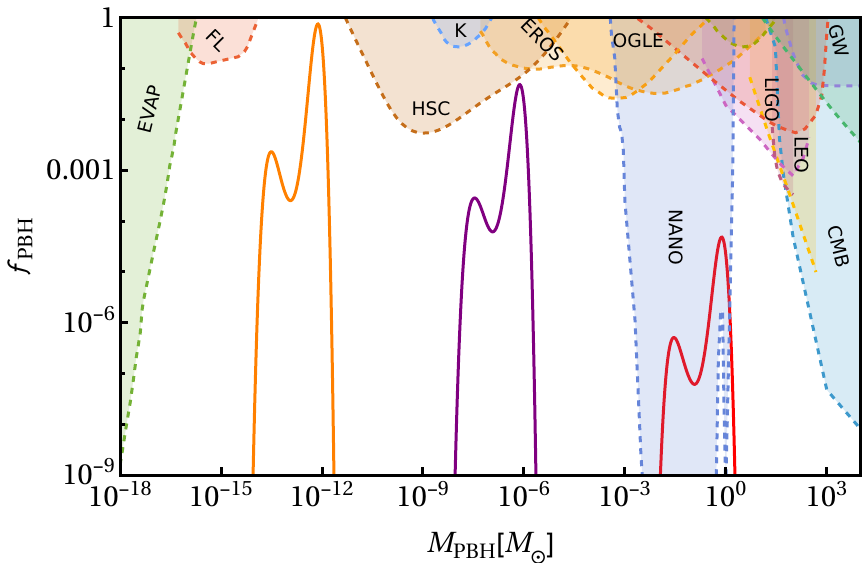} 
    \includegraphics[width=0.47\textwidth]{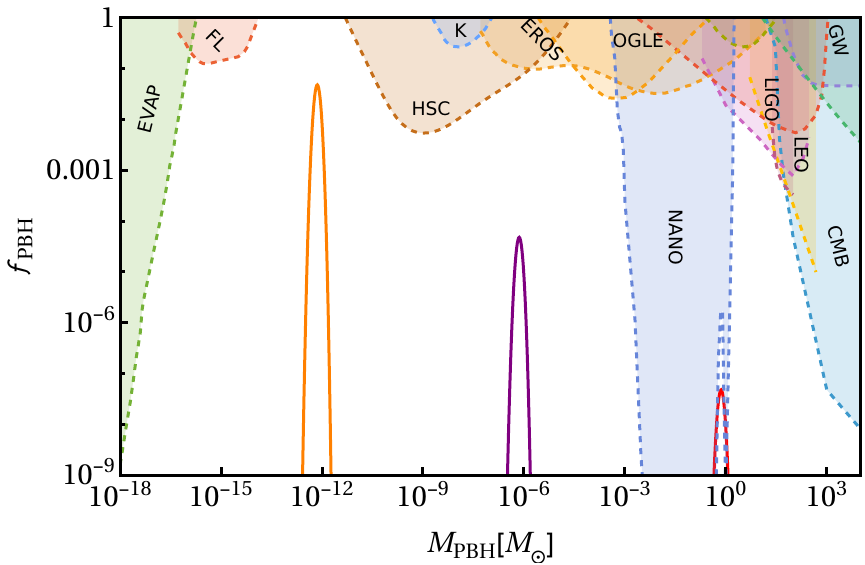} 
    \includegraphics[width=0.47\textwidth]{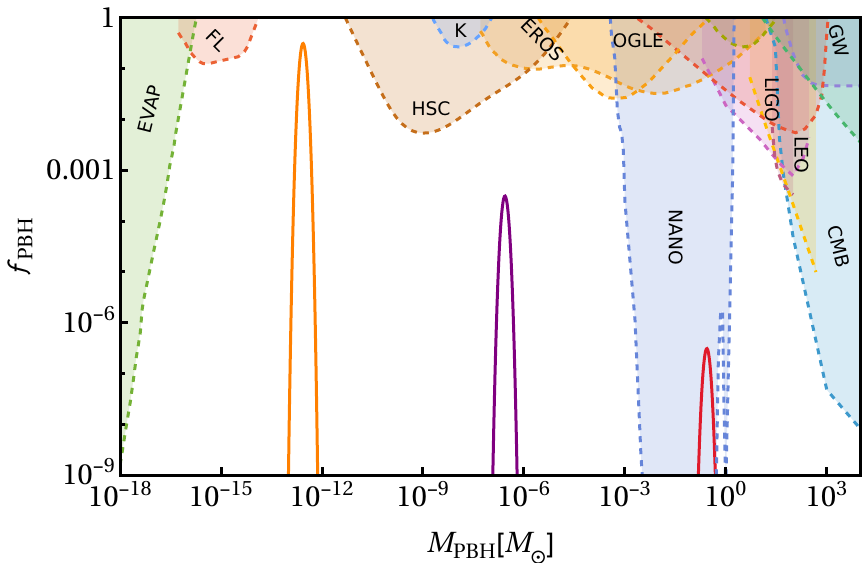}
    \caption{The abundance of primordial black holes (PBHs) is shown as a function of their mass, measured in solar mass units ($M_{\odot}$), for fixed values of $\epsilon_2$ across various combinations of $\eta_1$ (beginning of the USR phase) and $\eta_2$ (end of the USR phase). The {\color{red}Red} curves correspond to $\eta_2 = -3.33 \times 10^{-7}\ \text{Mpc}$, the {\color{purple}Purple} to $\eta_2 = -3.33 \times 10^{-10}\ \text{Mpc}$, and the {\color{Orange}Orange} to $\eta_2 = -3.33 \times 10^{-12}\ \text{Mpc}$. In the \textbf{first panel}, the \textbf{left plot} presents the PBH spectrum for $\epsilon_2 = -1$ with $\eta_1 = -2.86,\ -2.86 \times 10^{-3},\ -2.86 \times 10^{-5}\ \text{Mpc}$ (colored {\color{red}Red}, {\color{purple}Purple}, and {\color{Orange}Orange}, respectively), while the \textbf{right plot} shows results for $\epsilon_2 = -2$ using $\eta_1 = -9.09 \times 10^{-4},\ -9.09 \times 10^{-7},\ -9.09 \times 10^{-10}\ \text{Mpc}$ (same color scheme). The \textbf{second panel} displays the PBH abundance for $\epsilon_2 = -3$ on the \textbf{left}, with $\eta_1 = -6.33 \times 10^{-4},\ -6.33 \times 10^{-7},\ -6.33 \times 10^{-10}\ \text{Mpc}$, and for $\epsilon_2 = -6$ on the \textbf{right}, with $\eta_1 = -3.77 \times 10^{-5},\ -3.77 \times 10^{-8},\ -3.77 \times 10^{-11}\ \text{Mpc}$. Finally, in the \textbf{third panel}, the \textbf{left plot} illustrates the PBH spectrum for $\epsilon_2 = -8$ with $\eta_1 = -1.38 \times 10^{-5},\ -1.38 \times 10^{-8},\ -1.38 \times 10^{-11}\ \text{Mpc}$, and the \textbf{right plot} shows the case for $\epsilon_2 = -10$ using $\eta_1 = -9.22 \times 10^{-6},\ -9.22 \times 10^{-9},\ -9.22 \times 10^{-12}\ \text{Mpc}$, with colors following the same convention.}
    \label{fig: PBHplots}
\end{figure}

To relate PBH mass $M$ with scale $R$, we use the relation: $M=\gamma M_{H}$ where $M_H$ is the mass within the Hubble radius at horizon re-entry $(k = a H),$ and $\gamma$ is the efficiency of collapse. The scale-mass relation can be obtained as:
\begin{eqnarray}
  R(M)=\frac{2^{1/4}}{\gamma^{1/2}}\ \left(\frac{g_{\ast,k}}{g_{\ast, \rm eq}}\right)^{1/12}\ \left(\frac{1}{k_{\rm eq}}\right)\ \left(\frac{M}{M_{\rm eq}}\right)^{1/2}
\end{eqnarray}
where $k_{eq}$ is the wave number entering the horizon at matter-radiation equality, $M_{\rm eq} = 5.83\times 10^{47}$ Kg is the horizon mass at equality, and $g_{\ast,k}$ and  $g_{\ast,\rm eq}$ represents the number of relativistic degrees of freedom at the time of PBHs formation and radiation-matter equality, respectively. Expressed in solar masses $M_{\odot}$, the relation becomes:
\begin{eqnarray}
    R(M)=4.72\times 10^{-7}\ \left(\frac{\gamma}{0.2}\right)^{-1/2}\ \left(\frac{g_{*,k}}{g_{*,\rm eq}}\right)^{1/12}\ \left(\frac{M}{M_{\odot}}\right)^{1/2}\ \text{Mpc} 
\end{eqnarray}
Using this, we can finally define the present-day PBHs dark matter fraction $f_{\rm PBH}(M)$ as:
\begin{eqnarray}
    f_{\text{PBH}}(M)=2^{1/4}\ \gamma^{3/2}\ \beta(M)\ \left(\frac{\Omega_{m}h^{2}}{\Omega_{c}h^{2}}\right)\ \left(\frac{g_{*,k}}{g_{*,\rm eq}}\right)^{-1/4}\ \left(\frac{M}{M_{\rm eq}}\right)^{-1/2}
\end{eqnarray}
where $\Omega_{m}$ and $\Omega_{c}$ are the dimensionless matter and cold dark matter energy densities respectively.  and $H_{0} = 100 h \ \text{km} \,\text{sec}^{-1}\text{Mpc}^{-1}.$ For our analysis, we adopt the the parameter values as $\gamma = 0.2,\g_{\ast,k}=106.75,\ g_{\ast,\rm eq}=3.36, \, \Omega_{m}h^{2}=0.14$ and $\Omega_{c}h^{2}=0.12.$ Plugging these into the equation gives:
\begin{eqnarray}
    f_{\text{PBH}}(M)=\frac{\Omega_{\text{PBH}}}{\Omega_{\rm DM}}=\left(\frac{\gamma}{0.2}\right)^{3/2}\ \left(\frac{\beta(M)}{1.46\times10^{-8}}\right)\ \left(\frac{g_{*,k}}{g_{*,\rm eq}}\right)^{-1/4}\ \left(\frac{M}{M_{\rm eq}}\right)^{-1/2}
\end{eqnarray}
Today dark matter contributes $\sim 26\%$ of the total energy density of the present universe. However, whether PBHs account for all or only a fraction of this dark matter remains an open question. Several astrophysical observations and experiments impose stringent constraints on the abundance of PBHs, denoted by $f_{\text{PBH}}$, across a wide range of mass scales. These include limits from Hawking radiation \cite{Carr:2009jm}, Microlensing observations (Subaru HSC \cite{Niikura:2017zjd}, OGLE \cite{Niikura:2019kqi}, EROS-2 \cite{EROS-2:2006ryy}, and Kepler \cite{Griest:2013aaa}), CMB distortions \cite{Ali-Haimoud:2016mbv}, gravitational wave  observations \cite{LIGOScientific:2018glc, Carr:2020xqk, Chen:2019irf}, PTA \cite{Chen:2019xse}, Femtolensing \cite{Ezquiaga:2017fvi} and constraints from Dwarf galaxy dynamics \cite{Lu:2019ktw}. These observations collectively place stringent bounds on the viable parameter space for PBH formation. In Fig. \ref{fig: PBHplots}, we present the PBH abundance calculated from curvature perturbations generated at $k=10^6, 10^9$ and $10^{12}$ for different values of $\epsilon_{2}$ = $-1,-2,-3,-6,-8$ and $-10$. As expected, the results demonstrate that even for relatively mild values such as $\epsilon_{2}^{\rm usr} = -1$, it is possible to generate a significant PBH abundance. This clearly highlights that sub-horizon amplification of curvature perturbations alone can lead to PBH formation. Furthermore, the area under the Gaussian profiles indicates that the contribution from sub-horizon modes often exceeds that from super-horizon amplification.

This finding forms one of the key results of our work: {\bf sub-horizon growth of curvature perturbations, even in the absence of super-horizon enhancement, can play a dominant role in PBH formation.}

\section{Summary and Conclusions}\label{sec:conclu}

In this work, we propose a generalized mechanism for amplifying the curvature power spectrum during inflation, emphasizing the role of sub-horizon growth driven by the attenuation of the first slow-roll parameter $\epsilon_1$. Unlike the conventional requirement for ultra slow-roll (USR) conditions with $\epsilon_{2} \lesssim - 6$, we demonstrate that any negative value of the second slow-roll parameter $\epsilon_{2}$ can lead to significant enhancement of scalar perturbations --- provided it induces a strong suppression of $\epsilon_1$. This mechanism alone is sufficient to generate observable primordial gravitational waves (PGWs) and primordial black holes (PBHs), even in the absence of super-horizon amplification.

To illustrate this, we analyze a range of constant $\epsilon_{2}^{\rm usr}$ values and compute the resulting curvature power spectra, induced GW backgrounds, and PBH abundances. Notably, we show that even mild deviations from slow-roll, such as $\epsilon_{2}^{\rm usr} = -1$, can produce amplified perturbations that are compatible with or testable by current and upcoming observations, including PTA experiments like NANOGrav.

Crucially, the enhancement associated with USR dynamics arises naturally in inflationary models featuring localized structures in the potential --- such as bumps or inflection points --- which causes the inflaton field to decelerate. This deceleration dynamically induces a negative second slow-roll parameter, $\epsilon_{2}$ over a finite interval, thereby violating the standard slow-roll conditions and enabling significant amplification of curvature perturbations. Our analysis highlights that such behavior is not limited to specific potential shapes; even in the absence of explicit features, one can reconstruct the potential phenomenologically to achieve the required dynamics. This opens up a broader and more flexible framework for model building, allowing a wide class of inflationary scenarios to account for the production of primordial black holes (PBHs) and primordial gravitational waves (PGWs).

Furthermore, the implications of our results extend well beyond canonical single-field inflation. In modified theories of gravity, Galileon models or other non-canonical frameworks, the dynamics of scalar fields can be similarly engineered to induce sub-horizon amplification. This is especially relevant given that only a limited number of inflationary models are both theoretically well-motivated and consistent with particle physics principles. Hence, developing and constraining general amplification mechanisms --- like the one we explore --- is essential for gaining deeper insights into the physics of the early Universe.

Our central conclusion is clear: any inflationary model that permits a sub-horizon boost of curvature perturbations will necessarily yield an enhanced power spectrum. Additional features—such as inflection points or non-trivial interactions—can further enrich the model's phenomenology and observational signatures.

\section*{Acknowledgments}
DN is supported by the DST, Government of India through the DST-INSPIRE Faculty fellowship (04/2020/002142). The author thanks M. Sami, Mayukh Raj Gangopadhyay and Nilanjana Kumar for useful discussions and their valuable comments.

\providecommand{\href}[2]{#2}\begingroup\raggedright\endgroup

\end{document}